\documentclass[twocolumn,preprintnumbers,amsmath,amssymb,superscriptaddress]{revtex4-1}
\usepackage{graphicx}% Include figure files
\usepackage{float}
\usepackage{bm}% bold math
\usepackage{color}
\usepackage{ulem}
\begin{document}

%\title{Controlled generation of soliton gases in deep-water surface gravity waves}

\title{Nonlinear spectral synthesis of soliton gas in deep-water surface gravity waves}

\author{Pierre Suret}
\affiliation{Univ. Lille, CNRS, UMR 8523 - PhLAM -
  Physique des Lasers Atomes et Mol\'ecules, F-59 000 Lille, France}
\author{Alexey Tikan}
\affiliation{Univ. Lille, CNRS, UMR 8523 - PhLAM -
 Physique des Lasers Atomes et Mol\'ecules, F-59 000 Lille, France}
\author{F\'elicien Bonnefoy}
\affiliation{\'Ecole Centrale de Nantes, LHEEA, UMR 6598 CNRS, F-44 321 Nantes, France}
\author{Fran\c{c}ois Copie}
\affiliation{Univ. Lille, CNRS, UMR 8523 - PhLAM -
  Physique des Lasers Atomes et Mol\'ecules, F-59 000 Lille, France}
\author{Guillaume Ducrozet}
\affiliation{\'Ecole Centrale de Nantes, LHEEA, UMR 6598 CNRS, F-44 321 Nantes, France}
\author{Andrey Gelash}
\affiliation{Institute of Automation and Electrometry SB RAS, Novosibirsk 630090, Russia}
\affiliation{Skolkovo Institute of Science and Technology, Moscow 121205, Russia}
\author{Gaurav Prabhudesai}
\affiliation{Laboratoire de Physique de l'Ecole normale sup\'erieure, ENS,
Universit\'e PSL, CNRS, Sorbonne Universit\'e, Universit\'e Paris-Diderot, Paris, France}
\author{Guillaume Michel}
\affiliation{Sorbonne Universit\'e, CNRS, UMR 7190, Institut Jean Le Rond d’Alembert, F-75 005 Paris, France}
\author{Annette Cazaubiel}
\affiliation{Universit\'e de Paris, Universit\'e Paris Diderot, MSC, UMR 7057 CNRS, F-75 013 Paris, France}
\author{Eric Falcon}
\affiliation{Universit\'e de Paris, Universit\'e Paris Diderot,  MSC, UMR 7057 CNRS, F-75 013 Paris, France}
\author{Gennady El}
\affiliation{Department of Mathematics, Physics and Electrical Engineering, Northumbria University, Newcastle upon Tyne, NE1 8ST, United Kingdom}
\author{St\'ephane Randoux}
\email{stephane.randoux@univ-lille.fr}
\affiliation{Univ. Lille, CNRS, UMR 8523 - PhLAM -
  Physique des Lasers Atomes et Mol\'ecules, F-59 000 Lille, France}

\date{\today}% It is always \today, today,

\begin{abstract}
   
   Soliton gases represent large random soliton ensembles in physical systems
   that display integrable dynamics at the leading order. Despite significant theoretical
   developments and observational evidence of ubiquity of soliton gases in fluids and
   optical media  their controlled experimental realization has been missing.  We report
   the first controlled synthesis  of a dense soliton gas in deep-water surface gravity
   waves using the tools of nonlinear spectral theory (inverse scattering transform (IST))
   for the one-dimensional focusing nonlinear Schr\"odinger equation.
   The soliton gas is experimentally generated in a one-dimensional water tank 
   where we demonstrate that we can control and measure the density of states,
   i. e. the probability density function parametrizing
   the soliton gas in the IST spectral phase space.
   Nonlinear spectral analysis of the generated hydrodynamic soliton gas reveals that
   the density of states slowly changes under the influence of
   perturbative higher-order effects that break the integrability of the wave dynamics.  
  
\end{abstract}


\maketitle

Solitons are localized nonlinear waves
that have been studied in many areas of science over last
decades \cite{Remoissenet_book,Kartashov:11,Dauxoisbook}.
Solitons represent fundamental nonlinear modes of
physical systems described by a special class of wave equations of an
integrable nature \cite{Zabusky:65,Novikov_book,yang2010nonlinear}. These equations, 
like the Korteweg-de Vries (KdV) equation
or the one-dimensional nonlinear Schr\"odinger equation (1D-NLSE), are of 
significant physical importance since they describe
at the leading order the behavior of many systems in various fields of physics
such as water waves, matter waves or electromagnetic waves
\cite{Ablowitz:73,Remoissenet_book,Dauxoisbook,yang2010nonlinear,Trillo:16}. 

Nowadays the dynamics of soliton interaction is so well mastered that ordered
sets of optical solitons or their periodic generalizations, the so-called finite-gap potentials, are
synthesized  and manipulated to carry out
the transmission of information in fiber optics communication links 
\cite{Le:17,Turitsyn:17,le_nonlinear_2014,kamalian_signal_2018}. 
On the other hand, the question of collective dynamics of {\it large random} 
soliton ensembles represents a subject of active research in
statistical mechanics and in nonlinear physics,  
most notably in the contexts of ocean wave dynamics and
nonlinear optics, see e. g. ref. \cite{Osborne:80,Osborne:95,Onorato:01,Onorato:13,Pelinovsky_book,Onorato:05,Hassaini:17,Koussaifi:18,Randoux:14,Bromberg:10,SotoCrespo:16,Dudley:14,Kraych:19}. 

The concept of  soliton gas (SG)  as a large  ensemble of solitons
randomly distributed in space and elastically interacting with each
other originates from the work of Zakharov \cite{Zakharov:71}, who
introduced kinetic equation for  a non-equilibrium {\it diluted} gas
of weakly interacting solitons of the KdV equation. The
Zakharov's kinetic equation has been generalised to the case of a
dense SG in \cite{el_thermodynamic_2003} (KdV) and in
\cite{GEl:05, GEl:19} (focusing NLS).  Each soliton in a gas living on the infinite line $x$
is characterised by a discrete eigenvalue $\lambda_i$ of the spectrum of
the linear operator  associated with the  integrable evolution
equation within the inverse scattering transform (IST) formalism.  The
fundamental property of integrable dynamics is the preservation of the
soliton spectrum under evolution. The central concept in SG theory
is the density of states (DOS) \cite{lifshits_introduction_1988}
which represents the distribution $u(\lambda,x,t)$  over the spectral eigenvalues, so
that $u d \lambda dx$ is the number of soliton states found at time
$t$ in the element of the phase space $[\lambda, \lambda + d \lambda]\times [x,  x+ dx]$. 
The isospectrality of integrable dynamics results in the continuity
equation $u_t+(us)_x=0$ for the DOS evolution in a  spatially
nonhomogeneous (non-equilibrium) SG. The transport velocity
$s(\lambda,x,t)$ in the DOS continuity equation is different from the
free soliton velocity due to position/phase shifts in pairwise soliton
collisions, resulting in a non-local equation of state
$s=\mathcal{F}[u]$, relating the transport velocity with the DOS
\cite{GEl:05, GEl:19}. Interestingly, the SG  kinetic
equation has recently attracted much attention in the
context of generalized hydrodynamics  for quantum many-body integrable
systems, see  \cite{Doyon:18, doyon_geometric_2018, vu_equations_2019}
and references therein.

\begin{figure*}[!t]
  \includegraphics[width=1\textwidth]{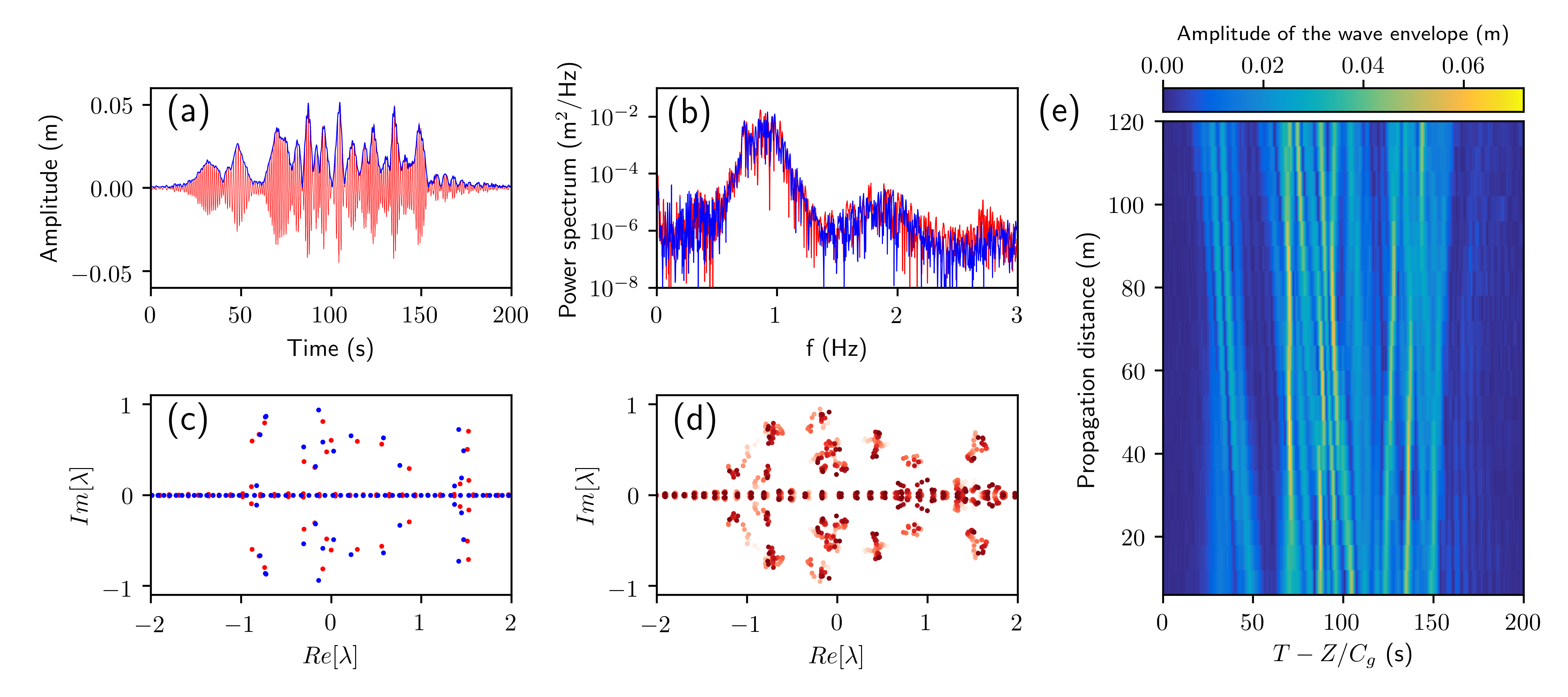}
  \caption{Ensemble of $N=16$ solitons propagating in the 1D water tank.
    (a) Water elevation (red line) and modulus of the wave envelope measured
    at $Z_1=6$ m, close to the wavemaker. (b) Fourier power spectra
    of wave elevation at $Z_1=6$ m (blue line) and at $Z_{20}=120$ m (red line).
    (c) Blue points represent the discrete IST spectrum of the numerically-generated
    N-SS $\psi_{16}(x,t=0)$ and red points represent the discrete IST spectrum
    measured at $Z_1=6$ m by using the signal plotted in (a). (d) Space evolution
    of the discrete IST spectra along the tank from $Z_1=6$ m (light red) to
    $Z_{20}=120$ m (dark red). (e) Space-time evolution of modulus of the
    wave envelope recorded by the $20$ gauges regularly spaced along the tank. 
    Physical parameters characterizing the experiment are $f_0=0.9$ Hz,
  $k_0=3.26$ m$^{-1}$, $\alpha=0.895$, $L_{NL}=210$ m ($\langle |A_0(T)|^2 \rangle=1.53 \, \times 10^{-4}$ m$^{2}$).} 
\end{figure*}

Despite various developments of SG theory (see  e.g. \cite{meiss_drift-wave_1982, fratalocchi_time-reversal_2011, GEl:11,dutykh_numerical_2014,  carbone_macroscopic_2016, shurgalina_nonlinear_2016, girotti_rigorous_2018, kachulin_soliton_2020} )   and the existence of
an unambiguous characterization of SG through the concept of DOS, the
experimental/observational results in this area are quite
limited. Costa {\it et al} have reported in 2014 the observation of
random wavepackets in shallow water ocean waves that have been
analyzed  using numerical IST tools and interpreted as randomly
distributed solitons that might be associated with KdV SG
\cite{Costa:14}. In 2015 large ensembles of interacting and colliding
solitons have been observed in a levitating rectilinear water cylinder \cite{Perrard:15}.
In the recent experiments reported in
ref. \cite{Redor:19}, Redor {\it et al} have taken advantage of the
process of fission of a sinusoidal wave train to generate an ensemble
of bidirectional shallow water solitons in a $34$-m long flume.
The interplay between multiple solitons and
dispersive radiation has been analyzed by Fourier tranform
and the observed random soliton ensemble has been interpreted
as representing a SG. In optics, the SG terminology
has been used to describe experiments
where light pulses were synchronously injected in a passive optical fiber ring cavity \cite{Schwache:97}. 
Another recent experimental observation of complex nonlinear wave behavior
attributed to SG dynamics was reported in
\cite{marcucci_topological_2019} where the formation of an incoherent
optical field has been observed in the long-time evolution of a square
pulse in a focusing medium \cite{GEl:16}.
However, in the absence of quantitative macroscopic (spectral) characterization the 
identification of the observed random wavefields with SG
remains questionable.  To our knowledge, there is no
existing experiment where SG have been unambiguously identified using IST
and where the measurement and control of the DOS of the SG have been achieved.

\begin{figure*}[!t]
  \includegraphics[width=1\textwidth]{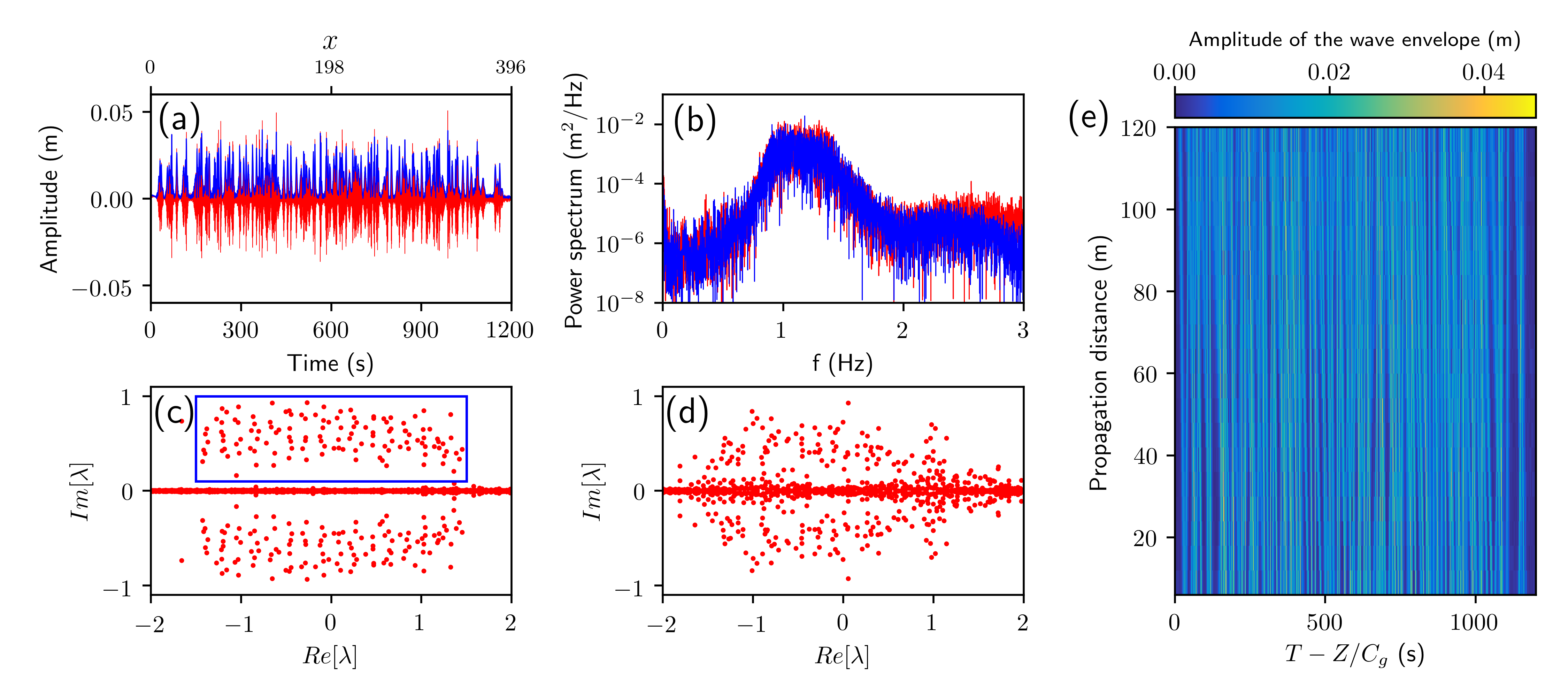}
  \caption{Gas of $N=128$ solitons propagating in the 1D water tank.
    (a) Water elevation (red line) and modulus of the wave envelope measured
    at $Z_1=6$ m, close to the wavemaker. (b) Fourier power spectra
    of wave elevation at $Z_1=6$ m (blue line) and at $Z_{20}=120$ m (red line).
    (c) Discrete IST spectrum measured at $Z_1=6$ m.
    (d) Discrete IST spectrum measured at $Z_{20}=120$ m.      
    (e) Space-time evolution of modulus of the
    wave envelope recorded by the $20$ gauges regularly spaced along the tank. 
    Physical parameters characterizing the experiment are $f_0=1.15$ Hz,
  $k_0=5.32$ m$^{-1}$, $\alpha=0.936$, $L_{NL}=45$ m ($\langle |A_0(T)|^2 \rangle=1.58 \,\times 10^{-4}$ m$^{2}$).
}
\end{figure*}

In this Letter, we report experiments fully based on the IST method where
we generate and observe the evolution of hydrodynamic deep-water dense soliton
gases. We take advantage of the recently developed methodology for the
effective numerical construction of the so-called $N$-soliton solutions of
the focusing 1D-NLSE with $N$ large (ref. \cite{Gelash:18}),
to create  an incoherent wavefield having a dominant and controlled 
solitonic content characterized by a measurable DOS. We show that the generated SG may undergo some
complex space-time evolution while the discrete IST spectrum is found to be
nearly conserved, albeit being perturbed by higher-order effects.
%We note that the IST based methods of the synthesis   of coherent (vanishing or periodic) signals  have been recently developed in \cite{le_nonlinear_2014, kamalian_signal_2018} in the context of optical telecommunications systems. Our method is principally designed to  create an {\it incoherent}  $N$-soliton ensemble that models a dense soliton gas with a given DOS. 

Our experiments were performed in a wave flume $148$ m long, $5$ m wide and $3$ m deep.
Unidirectional waves are generated at one end with a computer assisted flap-type wavemaker
and the flume is equipped with an absorbing device strongly reducing wave reflection at the
opposite end. As in the experiments reported in ref. \cite{Bonnefoy:20}, the setup comprises $20$
equally spaced resistive wave gauges that
are installed along the basin at distances $Z_j=6+(j-1)6$ m, $j=1,2,...20$
from the wavemaker located at $z=0$ m. This provides an effective measuring range
of $114$ m.

In our experiment, the water elevation at the wavemaker reads
$\eta(Z=0,T)=Re \left[ A_0(T)e^{i\omega_0 T} \right] $, where $\omega_0 = 2 \pi f_0$ is
the angular frequency of the carrier wave. $A_0(T)$ represents the complex envelope of the
initial condition. 
Our experiments are performed in the deep-water regime, and they are
designed in such a way that the observed dynamics
is described at leading order by the focusing 1D-NLSE
\begin{equation}\label{nlse_phys}
 \frac{\partial A}{\partial Z} + \frac{1}{C_g} \frac{\partial A}{\partial T}=  i \frac{k_0}{\omega_0^2}
\frac{\partial^2 A}{\partial T^2} + i  \alpha k_0^3 |A|^2 A,  
\end{equation}
where $A(Z,T)$ represents the complex envelope of the water wave
that changes in space $Z$ and in time $T$ \cite{osborne2010nonlinear}.
$k_0$ represents the wavenumber of the propagating wave ($\eta(Z,T)=Re \left[ A(Z,T)e^{i( \omega_0 T - k_0 Z)} \right]$),
which is linked to $\omega_0$ according to the deep
water dispersion relation $\omega_0^2=k_0 g$, where $g$ is the gravity acceleration.
$C_g=g/(2\omega_0)$ represents the group velocity of the wavepackets and
$\alpha$ is a dimensionless term describing the small finite-depth correction
to the cubic nonlinearity \cite{Bonnefoy:20}. 

The first important step of the experiment consists in generating an initial
condition $A_0(T)$ in the form of a random wavefield having a pure solitonic
content. To achieve this, we move to the ``IST-friendly'' canonical dimensionless form of the 1D-NLSE 
\begin{equation}\label{nlse}
i \frac{\partial \psi}{\partial t} + 
\frac{1}{2} \frac{\partial^2 \psi}{\partial x^2} + |\psi|^2 \psi=0,
\end{equation}
where $\psi(x,t)$ represents the normalized complex envelope of the water
wave.
Connection between physical  variables of Eq. (\ref{nlse_phys})
and  dimensionless variables in Eq. (\ref{nlse}) are given by $t=Z/L_{NL}$, $x=(T-Z/C_g)\sqrt{g/(2 L_{NL})}$
with the nonlinear length being defined as $L_{NL}=1/(\alpha k_0^3 \langle |A_0(T)|^2\rangle)$, where
the angle brackets denote average over time. 

The nonlinear wavefield $\psi(x, t)$ satisfying Eq. (\ref{nlse}) can be characterized
by the so-called scattering data (the IST spectrum).  For {\it localized}, i.e. decaying
to zero as $|x| \rightarrow \infty$ wavefield the IST spectrum consists of a discrete part related
to the soliton content and a continuous
part related to the dispersive radiation. A special class of
solutions, the $N$-soliton solutions (N-SS's),
exhibit only a discrete spectrum consisting of $N$ complex-valued
eigenvalues $\lambda_n$, $n=1,..., N$ and $N$ complex
parameters $C_n=|C_n|e^{i \phi_n}$, called norming constants, defined for
each $\lambda_n$.
%We stress that, while $|C_n|$ are functions of $\lambda_n$, the soliton phases $\phi_n$ represent an independent set of parameters.
In all the experiments described below, the phases $\phi_n$
of the norming constants $C_n$ characterizing the generated
N-SS are randomly and uniformly distributed over $[0, 2\pi)$ while
their modulus $|C_n|$ are chosen to be equal to unity. As shown in
ref. \cite{Gelash:18,Gelash:19}, such $N$-soliton statistical ensemble is a good model for a homogeneous
dense SG. 

\begin{figure*}[!t]
  \includegraphics[width=1\textwidth]{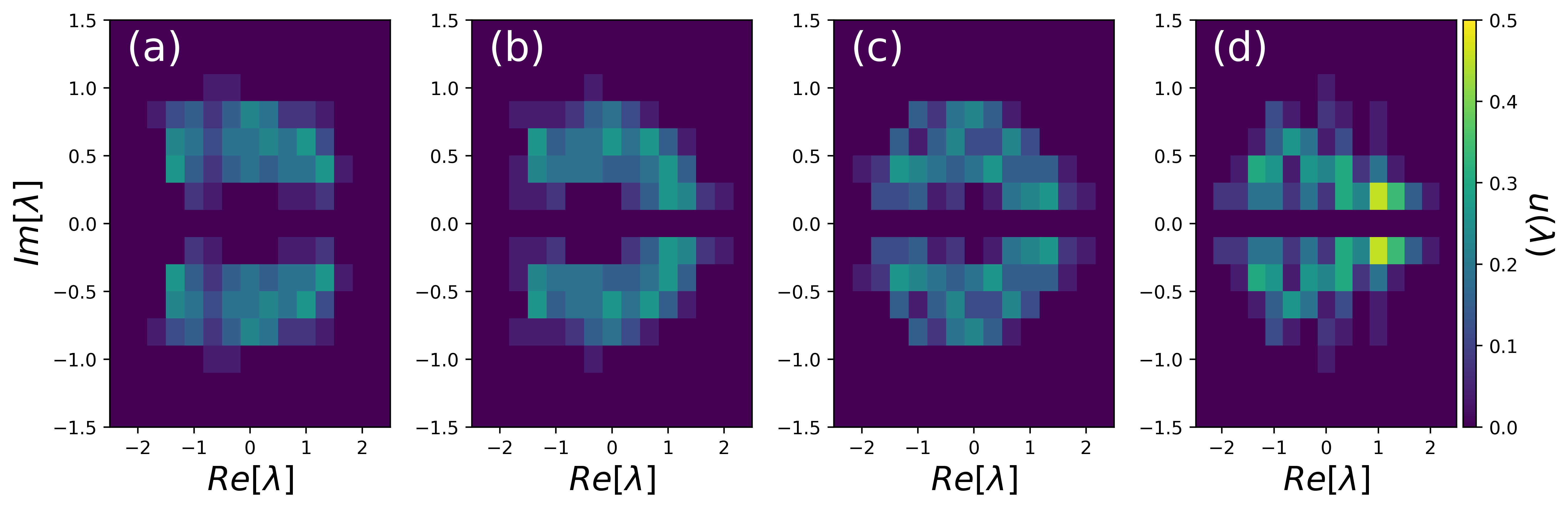}
  \caption{Statistical analysis of discrete IST spectra measured in Fig. 2
     showing the slow evolution of the
    DOS $u(\lambda)$ (the probability density function of the discrete IST eigenvalues in the
    complex plane) as a function of propagation distance in the water tank: 
    (a) $z_1=6$ m, (b) $z_3=18$ m, (c) $z_{10}=60$ m, (d) $z_{20}=120$ m.}
\end{figure*}

In our first experimental run, we used numerical methods described in ref. \cite{Gelash:18} to generate a N-SS of
Eq. (\ref{nlse}) (see also Supplemental Material \cite{note_suppl}), hereafter denoted $\psi_{16}(x,t)$,  with $N=16$
eigenvalues chosen arbitrarily within some domain of the complex spectral plane,
as shown with blue points in Fig. 1(c).
A relatively small number of solitons in this random soliton ensemble prevents its proper macroscopic spectral characterisation and the identification with SG. However, it is important as a first step in our experiment to establish a robust protocol for the generation of random soliton ensembles in a spectrally controlled way.

After some appropriate scaling, the generated dimensionless
wavefield $\psi_{16}(x,t=0)$ is converted into the physical complex envelope $A_{16}(Z=0,T)=A_0(T)$ of the
initial condition which is generated by the wavemaker. Fig. 1(a) shows the water elevation
measured at $Z_1=6$ m together with the modulus of the envelope $|A_{16}(Z_1,t)|$ computed
using standard Hilbert transform techniques \cite{osborne2010nonlinear}. The generated wavefield
with pure solitonic content spreads over approximately $140$ s and exhibits large  amplitude
fluctuations due to the random phase distribution. Fig. 1(c) shows the discrete
IST spectrum that is computed from the signal recorded by
the first gauge and plotted in Fig. 1(a). 
The measured eigenvalues plotted in red points in Fig. 1(c)
are close to the discrete eigenvalues (blue points) that we have selected to build 
$\psi_{16}(x,t=0)$, the N-SS under consideration. This demonstrates that the process of generation
of the N-SS solution is well controlled in our experiments.

As shown in Fig. 1(e), the space-time evolution of the generated wavepacket measured
with $20$ gauges distributed along the tank reveals complex dynamics with multiple
interacting coherent structures. At the same time, no significant broadening of the Fourier
power spectrum of the wavefield is observed between $Z_1=6$ m and $Z_{20}=120$ m,
as shown in Fig. 1(b). 
Despite the apparent complexity of the observed wave evolution, the
measured discrete IST spectra, compiled and superimposed in Fig. 1(d),
are nearly conserved over the whole propagation distance.

The fact that the isospectrality condition perfectly fullfilled in
a numerical simulation of the 1D-NLSE (see Supplemental Material \cite{note_suppl})
is not exactly verified in the experiment
arises from  perturbative higher-order effects that break the integrability
of the wave dynamics \cite{Chekhovskoy:19,Randoux:18,Bonnefoy:20},
see Supplemental Material showing numerical simulations revealing the trajectories
followed by eigenvalues in the complex spectral
plane under the influence of higher-order effects \cite{note_suppl}. In addition, 
the positions of the eigenvalues in Fig. 1(d)
are also perturbed because of measurement inaccurracies. Nevertheless, the results of
nonlinear spectral analysis reported in Fig. 1(d) show that the dynamical features observed for
the wavefield composed of $16$ solitons are nearly integrable.

We now take advantage of the above method of the controlled generation of multiple-soliton,
random phase solutions of the 1D-NLSE,  to generate a random $N$-soliton ensemble that can
be identified as SG. It is clear that to achieve that, the number of solitons $N$
should be sufficiently large.  Fig. 2 shows the dynamical and spectral
features characterizing the experimental evolution of an ensemble of $N=128$ solitons with random spectral (IST)
characteristics.
The important difference with the first example is that, due to a large number of solitons generated,
we are now able to characterize the  soliton ensemble by a DOS $u(\lambda)$, see Fig. 3.
Specifically, we generate a SG with eigenvalues $\lambda_i \in \mathbb{C}$ distributed
nearly uniformly on a rectangle in the upper half-plane of the complex IST spectral plane
(and the c.c. rectangle in the lower half plane) and the DOS $u(\lambda)=u_0$ being nearly constant
within the rectangle, see Fig. 2(c). 

Fig. 2(a) shows that the generated SG has the form of a random wavefield spreading over
$\Delta T=1200$ s which corresponds to a range $\Delta x =396$ in the dimensionless variables
of Eq. (\ref{nlse}).
Clearly the generated SG does not represent a {\it diluted} SG composed of
isolated and weakly interacting solitons but rather a {\it dense} SG which cannot be represented 
as superposition of individual solitons. Fig. 2(b) shows that the propagation of the generated SG 
is not accompanied by any significant broadening of Fourier power spectrum. 

Fig. 2(c) shows the discrete IST spectrum of the wavefield measured at $Z_1=6$ m, close
to the wavemaker. A set of $N=128$ eigenvalues is now measured within a rectangle in the upper complex
plane. Similarly to the features reported in Fig. 1, the perturbative higher-order effects
influence the observed dynamics and the discrete spectrum measured at $Z_{20}=120$ m is
not identical to the one measured at $Z_1$, see Fig. 2(d). Even though the isospectrality condition characterizing
a purely integrable dynamics is not exactly satisfied in our experiment, the measured discrete spectrum remains
confined to a well-defined region of the complex plane. Moreover, the large number of eigenvalues
distributed with some density within this limited region of the complex plane justifies the introduction of
a {\it statistical description of the spectral (IST) data}, which represents the key point for
the analysis of the observed wavefield in the framework of the SG theory. 

In the context of  the 1D-NLSE \eqref{nlse}  the DOS $u(\lambda)$, where $\lambda=\beta + i \gamma$,
represents the density of soliton states in the phase space i.e. $u d\beta d\gamma d x$ is the number of
solitons contained in a portion of  SG with the complex spectral parameter
$\lambda \in [\beta, \beta+d\beta] \times [\gamma, \gamma +d\gamma]$ over the space interval
$[x, x+dx]$  at time $t$ (corresponding to the position $Z$ in the tank).
Considering that the generated SG is homogeneous in space, the DOS represents the
probability density function of the complex-valued discrete eigenvalues normalised in
such a way that $ \int_{-\infty}^{+\infty} d\beta \int_{0}^{+\infty} d \gamma \, u(\lambda)  =N/\Delta x$,
where $N$ represents the number of eigenvalues found in the upper complex plane
and $\Delta x$ represents the spatial extent of the gas. 
Fig. 3 displays the normalized DOS experimentally measured at different propagation distances
in the water tank. We observe  slow evolution of the DOS along the tank which is
not due to gas' nonhomegeneity but mainly originates from the presence
of perturbative higher-order effects,
see Supplemental Material including numerical simulations 
revealing an evolution of the DOS similar to the one illustrated in Fig. 3 \cite{note_suppl}. Experimental results
reported in Fig. 3 suggest that the incorporation of higher-order perturbative physical effects
in the theory of SG represents a theoretical question of significant interest. 

In this Letter, we have reported hydrodynamic experiments demonstrating that
a controlled synthesis of a dense SG can be achieved in deep-water surface gravity waves.
We show that the generated SG is characterized by a measurable spectral DOS,
which provides an essential first step towards experimental verification of the kinetic theory of SGs.
We hope that our work will stimulate new experimental
and theoretical research in the fields of
statistical mechanics and nonlinear random waves. 

%\cite{Zakharov:72,Randoux:16a,Goullet:11,Dommermuth_1987,West_1987,Ducrozet_2012,CODE,Ducrozet_2012,Bonnefoy_2010}

\begin{acknowledgments}
This work has been partially supported  by the Agence Nationale de la
Recherche  through the LABEX CEMPI project (ANR-11-LABX-0007),   the
Ministry of Higher Education and Research, Hauts de France council and
European Regional Development  Fund (ERDF) through the Nord-Pas de
Calais Regional Research Council and the European Regional Development
Fund (ERDF) through the Contrat de Projets Etat-R\'egion (CPER
Photonics for Society P4S). The work of GE was partially supported by
EPSRC grant EP/R00515X/2. The work of FB, GD, GP,
GM, AC and EF was supported by the French National Research Agency
(ANR DYSTURB Project No. ANR-17-CE30-0004). EF thanks partial support
from the Simons Foundation/MPS No 651463. The work on the
construction of multisoliton ensembles was supported by
the Russian Science Foundation (Grant No. 19-72-30028
to A. G.). Simulations were partially performed at the
Novosibirsk Supercomputer Center (NSU).
\end{acknowledgments}

%\appendix

%\section{Appendixes}

%\begin{verbatim}
%\appendix
%\section{}
%\end{verbatim}
%will produce an appendix heading that says ``APPENDIX A'' and

%\bibliographystyle{apsrev4-1}
%\bibliography{bib_spectral,bib_book,bib_roguewaves}

%\bibliography{SG}

\begin{thebibliography}{61}%
\makeatletter
\providecommand \@ifxundefined [1]{%
 \@ifx{#1\undefined}
}%
\providecommand \@ifnum [1]{%
 \ifnum #1\expandafter \@firstoftwo
 \else \expandafter \@secondoftwo
 \fi
}%
\providecommand \@ifx [1]{%
 \ifx #1\expandafter \@firstoftwo
 \else \expandafter \@secondoftwo
 \fi
}%
\providecommand \natexlab [1]{#1}%
\providecommand \enquote  [1]{``#1''}%
\providecommand \bibnamefont  [1]{#1}%
\providecommand \bibfnamefont [1]{#1}%
\providecommand \citenamefont [1]{#1}%
\providecommand \href@noop [0]{\@secondoftwo}%
\providecommand \href [0]{\begingroup \@sanitize@url \@href}%
\providecommand \@href[1]{\@@startlink{#1}\@@href}%
\providecommand \@@href[1]{\endgroup#1\@@endlink}%
\providecommand \@sanitize@url [0]{\catcode `\\12\catcode `\$12\catcode
  `\&12\catcode `\#12\catcode `\^12\catcode `\_12\catcode `\%12\relax}%
\providecommand \@@startlink[1]{}%
\providecommand \@@endlink[0]{}%
\providecommand \url  [0]{\begingroup\@sanitize@url \@url }%
\providecommand \@url [1]{\endgroup\@href {#1}{\urlprefix }}%
\providecommand \urlprefix  [0]{URL }%
\providecommand \Eprint [0]{\href }%
\providecommand \doibase [0]{http://dx.doi.org/}%
\providecommand \selectlanguage [0]{\@gobble}%
\providecommand \bibinfo  [0]{\@secondoftwo}%
\providecommand \bibfield  [0]{\@secondoftwo}%
\providecommand \translation [1]{[#1]}%
\providecommand \BibitemOpen [0]{}%
\providecommand \bibitemStop [0]{}%
\providecommand \bibitemNoStop [0]{.\EOS\space}%
\providecommand \EOS [0]{\spacefactor3000\relax}%
\providecommand \BibitemShut  [1]{\csname bibitem#1\endcsname}%
\let\auto@bib@innerbib\@empty
%</preamble>
\bibitem [{\citenamefont {Remoissenet}(1996)}]{Remoissenet_book}%
  \BibitemOpen
  \bibfield  {author} {\bibinfo {author} {\bibfnamefont {M.}~\bibnamefont
  {Remoissenet}},\ }\href@noop {} {\emph {\bibinfo {title} {{Waves called
  solitons: concepts and experiments; 2nd ed.}}}}\ (\bibinfo  {publisher}
  {Springer},\ \bibinfo {address} {Berlin},\ \bibinfo {year}
  {1996})\BibitemShut {NoStop}%
\bibitem [{\citenamefont {Kartashov}\ \emph {et~al.}(2011)\citenamefont
  {Kartashov}, \citenamefont {Malomed},\ and\ \citenamefont
  {Torner}}]{Kartashov:11}%
  \BibitemOpen
  \bibfield  {author} {\bibinfo {author} {\bibfnamefont {Y.~V.}\ \bibnamefont
  {Kartashov}}, \bibinfo {author} {\bibfnamefont {B.~A.}\ \bibnamefont
  {Malomed}}, \ and\ \bibinfo {author} {\bibfnamefont {L.}~\bibnamefont
  {Torner}},\ }\href {\doibase 10.1103/RevModPhys.83.247} {\bibfield  {journal}
  {\bibinfo  {journal} {Rev. Mod. Phys.}\ }\textbf {\bibinfo {volume} {83}},\
  \bibinfo {pages} {247} (\bibinfo {year} {2011})}\BibitemShut {NoStop}%
\bibitem [{\citenamefont {Dauxois}\ and\ \citenamefont
  {Peyrard}(2006)}]{Dauxoisbook}%
  \BibitemOpen
  \bibfield  {author} {\bibinfo {author} {\bibfnamefont {T.}~\bibnamefont
  {Dauxois}}\ and\ \bibinfo {author} {\bibfnamefont {M.}~\bibnamefont
  {Peyrard}},\ }\href@noop {} {\emph {\bibinfo {title} {Physics of Solitons}}}\
  (\bibinfo  {publisher} {Cambridge University Press, Cambridge, England},\
  \bibinfo {year} {2006})\BibitemShut {NoStop}%
\bibitem [{\citenamefont {Zabusky}\ and\ \citenamefont
  {Kruskal}(1965)}]{Zabusky:65}%
  \BibitemOpen
  \bibfield  {author} {\bibinfo {author} {\bibfnamefont {N.~J.}\ \bibnamefont
  {Zabusky}}\ and\ \bibinfo {author} {\bibfnamefont {M.~D.}\ \bibnamefont
  {Kruskal}},\ }\href {\doibase 10.1103/PhysRevLett.15.240} {\bibfield
  {journal} {\bibinfo  {journal} {Phys. Rev. Lett.}\ }\textbf {\bibinfo
  {volume} {15}},\ \bibinfo {pages} {240} (\bibinfo {year} {1965})}\BibitemShut
  {NoStop}%
\bibitem [{\citenamefont {Novikov}\ \emph {et~al.}(1984)\citenamefont
  {Novikov}, \citenamefont {Manakov}, \citenamefont {Pitaevskii},\ and\
  \citenamefont {Zakharov}}]{Novikov_book}%
  \BibitemOpen
  \bibfield  {author} {\bibinfo {author} {\bibfnamefont {S.~P.}\ \bibnamefont
  {Novikov}}, \bibinfo {author} {\bibfnamefont {S.~V.}\ \bibnamefont
  {Manakov}}, \bibinfo {author} {\bibfnamefont {L.~P.}\ \bibnamefont
  {Pitaevskii}}, \ and\ \bibinfo {author} {\bibfnamefont {V.~E.}\ \bibnamefont
  {Zakharov}},\ }\href@noop {} {\emph {\bibinfo {title} {{Theory of solitons:
  the inverse scattering method}}}}\ (\bibinfo  {publisher} {Springer Science
  Business Media},\ \bibinfo {year} {1984})\BibitemShut {NoStop}%
\bibitem [{\citenamefont {Yang}(2010)}]{yang2010nonlinear}%
  \BibitemOpen
  \bibfield  {author} {\bibinfo {author} {\bibfnamefont {J.}~\bibnamefont
  {Yang}},\ }\href@noop {} {\emph {\bibinfo {title} {Nonlinear Waves in
  Integrable and Non-integrable Systems}}},\ Mathematical Modeling and
  Computation\ (\bibinfo  {publisher} {Society for Industrial and Applied
  Mathematics},\ \bibinfo {year} {2010})\BibitemShut {NoStop}%
\bibitem [{\citenamefont {Ablowitz}\ \emph {et~al.}(1973)\citenamefont
  {Ablowitz}, \citenamefont {Kaup}, \citenamefont {Newell},\ and\ \citenamefont
  {Segur}}]{Ablowitz:73}%
  \BibitemOpen
  \bibfield  {author} {\bibinfo {author} {\bibfnamefont {M.~J.}\ \bibnamefont
  {Ablowitz}}, \bibinfo {author} {\bibfnamefont {D.~J.}\ \bibnamefont {Kaup}},
  \bibinfo {author} {\bibfnamefont {A.~C.}\ \bibnamefont {Newell}}, \ and\
  \bibinfo {author} {\bibfnamefont {H.}~\bibnamefont {Segur}},\ }\href
  {\doibase 10.1103/PhysRevLett.31.125} {\bibfield  {journal} {\bibinfo
  {journal} {Phys. Rev. Lett.}\ }\textbf {\bibinfo {volume} {31}},\ \bibinfo
  {pages} {125} (\bibinfo {year} {1973})}\BibitemShut {NoStop}%
\bibitem [{\citenamefont {Trillo}\ \emph {et~al.}(2016)\citenamefont {Trillo},
  \citenamefont {Deng}, \citenamefont {Biondini}, \citenamefont {Klein},
  \citenamefont {Clauss}, \citenamefont {Chabchoub},\ and\ \citenamefont
  {Onorato}}]{Trillo:16}%
  \BibitemOpen
  \bibfield  {author} {\bibinfo {author} {\bibfnamefont {S.}~\bibnamefont
  {Trillo}}, \bibinfo {author} {\bibfnamefont {G.}~\bibnamefont {Deng}},
  \bibinfo {author} {\bibfnamefont {G.}~\bibnamefont {Biondini}}, \bibinfo
  {author} {\bibfnamefont {M.}~\bibnamefont {Klein}}, \bibinfo {author}
  {\bibfnamefont {G.~F.}\ \bibnamefont {Clauss}}, \bibinfo {author}
  {\bibfnamefont {A.}~\bibnamefont {Chabchoub}}, \ and\ \bibinfo {author}
  {\bibfnamefont {M.}~\bibnamefont {Onorato}},\ }\href {\doibase
  10.1103/PhysRevLett.117.144102} {\bibfield  {journal} {\bibinfo  {journal}
  {Phys. Rev. Lett.}\ }\textbf {\bibinfo {volume} {117}},\ \bibinfo {pages}
  {144102} (\bibinfo {year} {2016})}\BibitemShut {NoStop}%
\bibitem [{\citenamefont {Le}\ \emph {et~al.}(2017)\citenamefont {Le},
  \citenamefont {Aref},\ and\ \citenamefont {Buelow}}]{Le:17}%
  \BibitemOpen
  \bibfield  {author} {\bibinfo {author} {\bibfnamefont {S.~T.}\ \bibnamefont
  {Le}}, \bibinfo {author} {\bibfnamefont {V.}~\bibnamefont {Aref}}, \ and\
  \bibinfo {author} {\bibfnamefont {H.}~\bibnamefont {Buelow}},\ }\href@noop {}
  {\bibfield  {journal} {\bibinfo  {journal} {Nat. Photon.}\ }\textbf {\bibinfo
  {volume} {11}},\ \bibinfo {pages} {570} (\bibinfo {year} {2017})}\BibitemShut
  {NoStop}%
\bibitem [{\citenamefont {Turitsyn}\ \emph {et~al.}(2017)\citenamefont
  {Turitsyn}, \citenamefont {Prilepsky}, \citenamefont {Le}, \citenamefont
  {Wahls}, \citenamefont {Frumin}, \citenamefont {Kamalian},\ and\
  \citenamefont {Derevyanko}}]{Turitsyn:17}%
  \BibitemOpen
  \bibfield  {author} {\bibinfo {author} {\bibfnamefont {S.~K.}\ \bibnamefont
  {Turitsyn}}, \bibinfo {author} {\bibfnamefont {J.~E.}\ \bibnamefont
  {Prilepsky}}, \bibinfo {author} {\bibfnamefont {S.~T.}\ \bibnamefont {Le}},
  \bibinfo {author} {\bibfnamefont {S.}~\bibnamefont {Wahls}}, \bibinfo
  {author} {\bibfnamefont {L.~L.}\ \bibnamefont {Frumin}}, \bibinfo {author}
  {\bibfnamefont {M.}~\bibnamefont {Kamalian}}, \ and\ \bibinfo {author}
  {\bibfnamefont {S.~A.}\ \bibnamefont {Derevyanko}},\ }\href@noop {}
  {\bibfield  {journal} {\bibinfo  {journal} {Optica}\ }\textbf {\bibinfo
  {volume} {4}},\ \bibinfo {pages} {307} (\bibinfo {year} {2017})}\BibitemShut
  {NoStop}%
\bibitem [{\citenamefont {Le}\ \emph {et~al.}(2014)\citenamefont {Le},
  \citenamefont {Prilepsky},\ and\ \citenamefont
  {Turitsyn}}]{le_nonlinear_2014}%
  \BibitemOpen
  \bibfield  {author} {\bibinfo {author} {\bibfnamefont {S.~T.}\ \bibnamefont
  {Le}}, \bibinfo {author} {\bibfnamefont {J.~E.}\ \bibnamefont {Prilepsky}}, \
  and\ \bibinfo {author} {\bibfnamefont {S.~K.}\ \bibnamefont {Turitsyn}},\
  }\href {\doibase 10.1364/OE.22.026720} {\bibfield  {journal} {\bibinfo
  {journal} {Optics Express}\ }\textbf {\bibinfo {volume} {22}},\ \bibinfo
  {pages} {26720} (\bibinfo {year} {2014})}\BibitemShut {NoStop}%
\bibitem [{\citenamefont {Kamalian}\ \emph {et~al.}(2018)\citenamefont
  {Kamalian}, \citenamefont {Vasylchenkova}, \citenamefont {Shepelsky},
  \citenamefont {Prilepsky},\ and\ \citenamefont
  {Turitsyn}}]{kamalian_signal_2018}%
  \BibitemOpen
  \bibfield  {author} {\bibinfo {author} {\bibfnamefont {M.}~\bibnamefont
  {Kamalian}}, \bibinfo {author} {\bibfnamefont {A.}~\bibnamefont
  {Vasylchenkova}}, \bibinfo {author} {\bibfnamefont {D.}~\bibnamefont
  {Shepelsky}}, \bibinfo {author} {\bibfnamefont {J.~E.}\ \bibnamefont
  {Prilepsky}}, \ and\ \bibinfo {author} {\bibfnamefont {S.~K.}\ \bibnamefont
  {Turitsyn}},\ }\href {\doibase 10.1109/JLT.2018.2877103} {\bibfield
  {journal} {\bibinfo  {journal} {Journal of Lightwave Technology}\ }\textbf
  {\bibinfo {volume} {36}},\ \bibinfo {pages} {5714} (\bibinfo {year}
  {2018})}\BibitemShut {NoStop}%
\bibitem [{\citenamefont {Osborne}\ and\ \citenamefont
  {Burch}(1980)}]{Osborne:80}%
  \BibitemOpen
  \bibfield  {author} {\bibinfo {author} {\bibfnamefont {A.~R.}\ \bibnamefont
  {Osborne}}\ and\ \bibinfo {author} {\bibfnamefont {T.~L.}\ \bibnamefont
  {Burch}},\ }\href@noop {} {\bibfield  {journal} {\bibinfo  {journal}
  {Science}\ }\textbf {\bibinfo {volume} {208}},\ \bibinfo {pages} {451}
  (\bibinfo {year} {1980})}\BibitemShut {NoStop}%
\bibitem [{\citenamefont {Osborne}(1995)}]{Osborne:95}%
  \BibitemOpen
  \bibfield  {author} {\bibinfo {author} {\bibfnamefont {A.~R.}\ \bibnamefont
  {Osborne}},\ }\href {\doibase 10.1103/PhysRevE.52.1105} {\bibfield  {journal}
  {\bibinfo  {journal} {Phys. Rev. E}\ }\textbf {\bibinfo {volume} {52}},\
  \bibinfo {pages} {1105} (\bibinfo {year} {1995})}\BibitemShut {NoStop}%
\bibitem [{\citenamefont {Onorato}\ \emph {et~al.}(2001)\citenamefont
  {Onorato}, \citenamefont {Osborne}, \citenamefont {Serio},\ and\
  \citenamefont {Bertone}}]{Onorato:01}%
  \BibitemOpen
  \bibfield  {author} {\bibinfo {author} {\bibfnamefont {M.}~\bibnamefont
  {Onorato}}, \bibinfo {author} {\bibfnamefont {A.~R.}\ \bibnamefont
  {Osborne}}, \bibinfo {author} {\bibfnamefont {M.}~\bibnamefont {Serio}}, \
  and\ \bibinfo {author} {\bibfnamefont {S.}~\bibnamefont {Bertone}},\ }\href
  {\doibase 10.1103/PhysRevLett.86.5831} {\bibfield  {journal} {\bibinfo
  {journal} {Phys. Rev. Lett.}\ }\textbf {\bibinfo {volume} {86}},\ \bibinfo
  {pages} {5831} (\bibinfo {year} {2001})}\BibitemShut {NoStop}%
\bibitem [{\citenamefont {Onorato}\ \emph {et~al.}(2013)\citenamefont
  {Onorato}, \citenamefont {Residori}, \citenamefont {Bortolozzo},
  \citenamefont {Montina},\ and\ \citenamefont {Arecchi}}]{Onorato:13}%
  \BibitemOpen
  \bibfield  {author} {\bibinfo {author} {\bibfnamefont {M.}~\bibnamefont
  {Onorato}}, \bibinfo {author} {\bibfnamefont {S.}~\bibnamefont {Residori}},
  \bibinfo {author} {\bibfnamefont {U.}~\bibnamefont {Bortolozzo}}, \bibinfo
  {author} {\bibfnamefont {A.}~\bibnamefont {Montina}}, \ and\ \bibinfo
  {author} {\bibfnamefont {F.}~\bibnamefont {Arecchi}},\ }\href {\doibase
  http://dx.doi.org/10.1016/j.physrep.2013.03.001} {\bibfield  {journal}
  {\bibinfo  {journal} {Phys. Rep.}\ }\textbf {\bibinfo {volume} {528}},\
  \bibinfo {pages} {47 } (\bibinfo {year} {2013})}\BibitemShut {NoStop}%
\bibitem [{\citenamefont {Pelinovsky}\ \emph {et~al.}(2008)\citenamefont
  {Pelinovsky}, \citenamefont {Kharif} \emph {et~al.}}]{Pelinovsky_book}%
  \BibitemOpen
  \bibfield  {author} {\bibinfo {author} {\bibfnamefont {E.}~\bibnamefont
  {Pelinovsky}}, \bibinfo {author} {\bibfnamefont {C.}~\bibnamefont {Kharif}},
  \emph {et~al.},\ }\href@noop {} {\emph {\bibinfo {title} {Extreme ocean
  waves}}}\ (\bibinfo  {publisher} {Springer},\ \bibinfo {year}
  {2008})\BibitemShut {NoStop}%
\bibitem [{\citenamefont {Onorato}\ \emph {et~al.}(2005)\citenamefont
  {Onorato}, \citenamefont {Osborne}, \citenamefont {Serio},\ and\
  \citenamefont {Cavaleri}}]{Onorato:05}%
  \BibitemOpen
  \bibfield  {author} {\bibinfo {author} {\bibfnamefont {M.}~\bibnamefont
  {Onorato}}, \bibinfo {author} {\bibfnamefont {A.~R.}\ \bibnamefont
  {Osborne}}, \bibinfo {author} {\bibfnamefont {M.}~\bibnamefont {Serio}}, \
  and\ \bibinfo {author} {\bibfnamefont {L.}~\bibnamefont {Cavaleri}},\
  }\href@noop {} {\bibfield  {journal} {\bibinfo  {journal} {Physics of
  Fluids}\ }\textbf {\bibinfo {volume} {17}},\ \bibinfo {pages} {078101}
  (\bibinfo {year} {2005})}\BibitemShut {NoStop}%
\bibitem [{\citenamefont {Hassaini}\ and\ \citenamefont
  {Mordant}(2017)}]{Hassaini:17}%
  \BibitemOpen
  \bibfield  {author} {\bibinfo {author} {\bibfnamefont {R.}~\bibnamefont
  {Hassaini}}\ and\ \bibinfo {author} {\bibfnamefont {N.}~\bibnamefont
  {Mordant}},\ }\href {\doibase 10.1103/PhysRevFluids.2.094803} {\bibfield
  {journal} {\bibinfo  {journal} {Phys. Rev. Fluids}\ }\textbf {\bibinfo
  {volume} {2}},\ \bibinfo {pages} {094803} (\bibinfo {year}
  {2017})}\BibitemShut {NoStop}%
\bibitem [{\citenamefont {El~Koussaifi}\ \emph {et~al.}(2018)\citenamefont
  {El~Koussaifi}, \citenamefont {Tikan}, \citenamefont {Toffoli}, \citenamefont
  {Randoux}, \citenamefont {Suret},\ and\ \citenamefont
  {Onorato}}]{Koussaifi:18}%
  \BibitemOpen
  \bibfield  {author} {\bibinfo {author} {\bibfnamefont {R.}~\bibnamefont
  {El~Koussaifi}}, \bibinfo {author} {\bibfnamefont {A.}~\bibnamefont {Tikan}},
  \bibinfo {author} {\bibfnamefont {A.}~\bibnamefont {Toffoli}}, \bibinfo
  {author} {\bibfnamefont {S.}~\bibnamefont {Randoux}}, \bibinfo {author}
  {\bibfnamefont {P.}~\bibnamefont {Suret}}, \ and\ \bibinfo {author}
  {\bibfnamefont {M.}~\bibnamefont {Onorato}},\ }\href {\doibase
  10.1103/PhysRevE.97.012208} {\bibfield  {journal} {\bibinfo  {journal} {Phys.
  Rev. E}\ }\textbf {\bibinfo {volume} {97}},\ \bibinfo {pages} {012208}
  (\bibinfo {year} {2018})}\BibitemShut {NoStop}%
\bibitem [{\citenamefont {Randoux}\ \emph {et~al.}(2014)\citenamefont
  {Randoux}, \citenamefont {Walczak}, \citenamefont {Onorato},\ and\
  \citenamefont {Suret}}]{Randoux:14}%
  \BibitemOpen
  \bibfield  {author} {\bibinfo {author} {\bibfnamefont {S.}~\bibnamefont
  {Randoux}}, \bibinfo {author} {\bibfnamefont {P.}~\bibnamefont {Walczak}},
  \bibinfo {author} {\bibfnamefont {M.}~\bibnamefont {Onorato}}, \ and\
  \bibinfo {author} {\bibfnamefont {P.}~\bibnamefont {Suret}},\ }\href
  {\doibase 10.1103/PhysRevLett.113.113902} {\bibfield  {journal} {\bibinfo
  {journal} {Phys. Rev. Lett.}\ }\textbf {\bibinfo {volume} {113}},\ \bibinfo
  {pages} {113902} (\bibinfo {year} {2014})}\BibitemShut {NoStop}%
\bibitem [{\citenamefont {Bromberg}\ \emph {et~al.}(2010)\citenamefont
  {Bromberg}, \citenamefont {Lahini}, \citenamefont {Small},\ and\
  \citenamefont {Silberberg}}]{Bromberg:10}%
  \BibitemOpen
  \bibfield  {author} {\bibinfo {author} {\bibfnamefont {Y.}~\bibnamefont
  {Bromberg}}, \bibinfo {author} {\bibfnamefont {U.}~\bibnamefont {Lahini}},
  \bibinfo {author} {\bibfnamefont {E.}~\bibnamefont {Small}}, \ and\ \bibinfo
  {author} {\bibfnamefont {Y.}~\bibnamefont {Silberberg}},\ }\href@noop {}
  {\bibfield  {journal} {\bibinfo  {journal} {Nat. Photon.}\ }\textbf {\bibinfo
  {volume} {4}},\ \bibinfo {pages} {721} (\bibinfo {year} {2010})}\BibitemShut
  {NoStop}%
\bibitem [{\citenamefont {Soto-Crespo}\ \emph {et~al.}(2016)\citenamefont
  {Soto-Crespo}, \citenamefont {Devine},\ and\ \citenamefont
  {Akhmediev}}]{SotoCrespo:16}%
  \BibitemOpen
  \bibfield  {author} {\bibinfo {author} {\bibfnamefont {J.~M.}\ \bibnamefont
  {Soto-Crespo}}, \bibinfo {author} {\bibfnamefont {N.}~\bibnamefont {Devine}},
  \ and\ \bibinfo {author} {\bibfnamefont {N.}~\bibnamefont {Akhmediev}},\
  }\href {\doibase 10.1103/PhysRevLett.116.103901} {\bibfield  {journal}
  {\bibinfo  {journal} {Phys. Rev. Lett.}\ }\textbf {\bibinfo {volume} {116}},\
  \bibinfo {pages} {103901} (\bibinfo {year} {2016})}\BibitemShut {NoStop}%
\bibitem [{\citenamefont {Dudley}\ \emph {et~al.}(2014)\citenamefont {Dudley},
  \citenamefont {Dias}, \citenamefont {Erkintalo},\ and\ \citenamefont
  {Genty}}]{Dudley:14}%
  \BibitemOpen
  \bibfield  {author} {\bibinfo {author} {\bibfnamefont {J.~M.}\ \bibnamefont
  {Dudley}}, \bibinfo {author} {\bibfnamefont {F.}~\bibnamefont {Dias}},
  \bibinfo {author} {\bibfnamefont {M.}~\bibnamefont {Erkintalo}}, \ and\
  \bibinfo {author} {\bibfnamefont {G.}~\bibnamefont {Genty}},\ }\href@noop {}
  {\bibfield  {journal} {\bibinfo  {journal} {Nature Photonics}\ }\textbf
  {\bibinfo {volume} {8}},\ \bibinfo {pages} {755} (\bibinfo {year}
  {2014})}\BibitemShut {NoStop}%
\bibitem [{\citenamefont {Kraych}\ \emph {et~al.}(2019)\citenamefont {Kraych},
  \citenamefont {Agafontsev}, \citenamefont {Randoux},\ and\ \citenamefont
  {Suret}}]{Kraych:19}%
  \BibitemOpen
  \bibfield  {author} {\bibinfo {author} {\bibfnamefont {A.~E.}\ \bibnamefont
  {Kraych}}, \bibinfo {author} {\bibfnamefont {D.}~\bibnamefont {Agafontsev}},
  \bibinfo {author} {\bibfnamefont {S.}~\bibnamefont {Randoux}}, \ and\
  \bibinfo {author} {\bibfnamefont {P.}~\bibnamefont {Suret}},\ }\href
  {\doibase 10.1103/PhysRevLett.123.093902} {\bibfield  {journal} {\bibinfo
  {journal} {Phys. Rev. Lett.}\ }\textbf {\bibinfo {volume} {123}},\ \bibinfo
  {pages} {093902} (\bibinfo {year} {2019})}\BibitemShut {NoStop}%
\bibitem [{\citenamefont {Zakharov}(1971)}]{Zakharov:71}%
  \BibitemOpen
  \bibfield  {author} {\bibinfo {author} {\bibfnamefont {V.~E.}\ \bibnamefont
  {Zakharov}},\ }\href@noop {} {\bibfield  {journal} {\bibinfo  {journal} {Sov.
  Phys.--JETP}\ }\textbf {\bibinfo {volume} {33}},\ \bibinfo {pages} {538}
  (\bibinfo {year} {1971})}\BibitemShut {NoStop}%
\bibitem [{\citenamefont {El}(2003)}]{el_thermodynamic_2003}%
  \BibitemOpen
  \bibfield  {author} {\bibinfo {author} {\bibfnamefont {G.}~\bibnamefont
  {El}},\ }\href {\doibase 10.1016/S0375-9601(03)00515-2} {\bibfield  {journal}
  {\bibinfo  {journal} {Physics Letters A}\ }\textbf {\bibinfo {volume}
  {311}},\ \bibinfo {pages} {374} (\bibinfo {year} {2003})}\BibitemShut
  {NoStop}%
\bibitem [{\citenamefont {El}\ and\ \citenamefont {Kamchatnov}(2005)}]{GEl:05}%
  \BibitemOpen
  \bibfield  {author} {\bibinfo {author} {\bibfnamefont {G.~A.}\ \bibnamefont
  {El}}\ and\ \bibinfo {author} {\bibfnamefont {A.~M.}\ \bibnamefont
  {Kamchatnov}},\ }\href {\doibase 10.1103/PhysRevLett.95.204101} {\bibfield
  {journal} {\bibinfo  {journal} {Phys. Rev. Lett.}\ }\textbf {\bibinfo
  {volume} {95}},\ \bibinfo {pages} {204101} (\bibinfo {year}
  {2005})}\BibitemShut {NoStop}%
\bibitem [{\citenamefont {El}\ and\ \citenamefont {Tovbis}(2020)}]{GEl:19}%
  \BibitemOpen
  \bibfield  {author} {\bibinfo {author} {\bibfnamefont {G.}~\bibnamefont
  {El}}\ and\ \bibinfo {author} {\bibfnamefont {A.}~\bibnamefont {Tovbis}},\
  }\href@noop {} {\bibfield  {journal} {\bibinfo  {journal} {Phys. Rev. E}\
  }\textbf {\bibinfo {volume} {101}},\ \bibinfo {pages} {052207} (\bibinfo
  {year} {2020})}\BibitemShut {NoStop}%
\bibitem [{\citenamefont {Lifshits}\ \emph {et~al.}(1988)\citenamefont
  {Lifshits}, \citenamefont {Gredeskul},\ and\ \citenamefont
  {Pastur}}]{lifshits_introduction_1988}%
  \BibitemOpen
  \bibfield  {author} {\bibinfo {author} {\bibfnamefont {I.}~\bibnamefont
  {Lifshits}}, \bibinfo {author} {\bibfnamefont {S.}~\bibnamefont {Gredeskul}},
  \ and\ \bibinfo {author} {\bibfnamefont {L.}~\bibnamefont {Pastur}},\
  }\href@noop {} {\emph {\bibinfo {title} {Introduction to the theory of
  disordered systems}}}\ (\bibinfo  {publisher} {Wiley},\ \bibinfo {year}
  {1988})\BibitemShut {NoStop}%
\bibitem [{\citenamefont {Doyon}\ \emph
  {et~al.}(2018{\natexlab{a}})\citenamefont {Doyon}, \citenamefont
  {Yoshimura},\ and\ \citenamefont {Caux}}]{Doyon:18}%
  \BibitemOpen
  \bibfield  {author} {\bibinfo {author} {\bibfnamefont {B.}~\bibnamefont
  {Doyon}}, \bibinfo {author} {\bibfnamefont {T.}~\bibnamefont {Yoshimura}}, \
  and\ \bibinfo {author} {\bibfnamefont {J.-S.}\ \bibnamefont {Caux}},\ }\href
  {\doibase 10.1103/PhysRevLett.120.045301} {\bibfield  {journal} {\bibinfo
  {journal} {Phys. Rev. Lett.}\ }\textbf {\bibinfo {volume} {120}},\ \bibinfo
  {pages} {045301} (\bibinfo {year} {2018}{\natexlab{a}})}\BibitemShut
  {NoStop}%
\bibitem [{\citenamefont {Doyon}\ \emph
  {et~al.}(2018{\natexlab{b}})\citenamefont {Doyon}, \citenamefont {Spohn},\
  and\ \citenamefont {Yoshimura}}]{doyon_geometric_2018}%
  \BibitemOpen
  \bibfield  {author} {\bibinfo {author} {\bibfnamefont {B.}~\bibnamefont
  {Doyon}}, \bibinfo {author} {\bibfnamefont {H.}~\bibnamefont {Spohn}}, \ and\
  \bibinfo {author} {\bibfnamefont {T.}~\bibnamefont {Yoshimura}},\ }\href@noop
  {} {\bibfield  {journal} {\bibinfo  {journal} {Nuclear Physics B}\ }\textbf
  {\bibinfo {volume} {926}},\ \bibinfo {pages} {570} (\bibinfo {year}
  {2018}{\natexlab{b}})}\BibitemShut {NoStop}%
\bibitem [{\citenamefont {Vu}\ and\ \citenamefont
  {Yoshimura}(2019)}]{vu_equations_2019}%
  \BibitemOpen
  \bibfield  {author} {\bibinfo {author} {\bibfnamefont {D.-L.}\ \bibnamefont
  {Vu}}\ and\ \bibinfo {author} {\bibfnamefont {T.}~\bibnamefont {Yoshimura}},\
  }\href@noop {} {\bibfield  {journal} {\bibinfo  {journal} {SciPost Physics}\
  }\textbf {\bibinfo {volume} {6}} (\bibinfo {year} {2019})}\BibitemShut
  {NoStop}%
\bibitem [{\citenamefont {Meiss}\ and\ \citenamefont
  {Horton~Jr}(1982)}]{meiss_drift-wave_1982}%
  \BibitemOpen
  \bibfield  {author} {\bibinfo {author} {\bibfnamefont {J.~D.}\ \bibnamefont
  {Meiss}}\ and\ \bibinfo {author} {\bibfnamefont {W.}~\bibnamefont
  {Horton~Jr}},\ }\href@noop {} {\bibfield  {journal} {\bibinfo  {journal}
  {Physical Review Letters}\ }\textbf {\bibinfo {volume} {48}},\ \bibinfo
  {pages} {1362} (\bibinfo {year} {1982})}\BibitemShut {NoStop}%
\bibitem [{\citenamefont {Fratalocchi}\ \emph {et~al.}(2011)\citenamefont
  {Fratalocchi}, \citenamefont {Armaroli},\ and\ \citenamefont
  {Trillo}}]{fratalocchi_time-reversal_2011}%
  \BibitemOpen
  \bibfield  {author} {\bibinfo {author} {\bibfnamefont {A.}~\bibnamefont
  {Fratalocchi}}, \bibinfo {author} {\bibfnamefont {A.}~\bibnamefont
  {Armaroli}}, \ and\ \bibinfo {author} {\bibfnamefont {S.}~\bibnamefont
  {Trillo}},\ }\href {\doibase 10.1103/PhysRevA.83.053846} {\bibfield
  {journal} {\bibinfo  {journal} {Physical Review A}\ }\textbf {\bibinfo
  {volume} {83}} (\bibinfo {year} {2011}),\
  10.1103/PhysRevA.83.053846}\BibitemShut {NoStop}%
\bibitem [{\citenamefont {El}\ \emph {et~al.}(2011)\citenamefont {El},
  \citenamefont {Kamchatnov}, \citenamefont {Pavlov},\ and\ \citenamefont
  {Zykov}}]{GEl:11}%
  \BibitemOpen
  \bibfield  {author} {\bibinfo {author} {\bibfnamefont {G.~A.}\ \bibnamefont
  {El}}, \bibinfo {author} {\bibfnamefont {A.~M.}\ \bibnamefont {Kamchatnov}},
  \bibinfo {author} {\bibfnamefont {M.~V.}\ \bibnamefont {Pavlov}}, \ and\
  \bibinfo {author} {\bibfnamefont {S.~A.}\ \bibnamefont {Zykov}},\ }\href
  {\doibase 10.1007/s00332-010-9080-z} {\bibfield  {journal} {\bibinfo
  {journal} {Journal of Nonlinear Science}\ }\textbf {\bibinfo {volume} {21}},\
  \bibinfo {pages} {151} (\bibinfo {year} {2011})}\BibitemShut {NoStop}%
\bibitem [{\citenamefont {Dutykh}\ and\ \citenamefont
  {Pelinovsky}(2014)}]{dutykh_numerical_2014}%
  \BibitemOpen
  \bibfield  {author} {\bibinfo {author} {\bibfnamefont {D.}~\bibnamefont
  {Dutykh}}\ and\ \bibinfo {author} {\bibfnamefont {E.}~\bibnamefont
  {Pelinovsky}},\ }\href {\doibase 10.1016/j.physleta.2014.09.008} {\bibfield
  {journal} {\bibinfo  {journal} {Physics Letters A}\ }\textbf {\bibinfo
  {volume} {378}},\ \bibinfo {pages} {3102} (\bibinfo {year}
  {2014})}\BibitemShut {NoStop}%
\bibitem [{\citenamefont {Carbone}\ \emph {et~al.}(2016)\citenamefont
  {Carbone}, \citenamefont {Dutykh},\ and\ \citenamefont
  {El}}]{carbone_macroscopic_2016}%
  \BibitemOpen
  \bibfield  {author} {\bibinfo {author} {\bibfnamefont {F.}~\bibnamefont
  {Carbone}}, \bibinfo {author} {\bibfnamefont {D.}~\bibnamefont {Dutykh}}, \
  and\ \bibinfo {author} {\bibfnamefont {G.~A.}\ \bibnamefont {El}},\
  }\href@noop {} {\bibfield  {journal} {\bibinfo  {journal} {EPL (Europhysics
  Letters)}\ }\textbf {\bibinfo {volume} {113}},\ \bibinfo {pages} {30003}
  (\bibinfo {year} {2016})}\BibitemShut {NoStop}%
\bibitem [{\citenamefont {Shurgalina}\ and\ \citenamefont
  {Pelinovsky}(2016)}]{shurgalina_nonlinear_2016}%
  \BibitemOpen
  \bibfield  {author} {\bibinfo {author} {\bibfnamefont {E.}~\bibnamefont
  {Shurgalina}}\ and\ \bibinfo {author} {\bibfnamefont {E.}~\bibnamefont
  {Pelinovsky}},\ }\href {\doibase 10.1016/j.physleta.2016.04.023} {\bibfield
  {journal} {\bibinfo  {journal} {Physics Letters A}\ }\textbf {\bibinfo
  {volume} {380}},\ \bibinfo {pages} {2049} (\bibinfo {year}
  {2016})}\BibitemShut {NoStop}%
\bibitem [{\citenamefont {Girotti}\ \emph {et~al.}(2018)\citenamefont
  {Girotti}, \citenamefont {Grava},\ and\ \citenamefont
  {McLaughlin}}]{girotti_rigorous_2018}%
  \BibitemOpen
  \bibfield  {author} {\bibinfo {author} {\bibfnamefont {M.}~\bibnamefont
  {Girotti}}, \bibinfo {author} {\bibfnamefont {T.}~\bibnamefont {Grava}}, \
  and\ \bibinfo {author} {\bibfnamefont {K.~D. T.-R.}\ \bibnamefont
  {McLaughlin}},\ }\href {http://arxiv.org/abs/1807.00608} {\bibfield
  {journal} {\bibinfo  {journal} {arXiv:1807.00608 [math-ph, physics:nlin]}\ }
  (\bibinfo {year} {2018})},\ \bibinfo {note} {arXiv: 1807.00608}\BibitemShut
  {NoStop}%
\bibitem [{\citenamefont {Kachulin}\ \emph {et~al.}(2020)\citenamefont
  {Kachulin}, \citenamefont {Dyachenko},\ and\ \citenamefont
  {Zakharov}}]{kachulin_soliton_2020}%
  \BibitemOpen
  \bibfield  {author} {\bibinfo {author} {\bibfnamefont {D.}~\bibnamefont
  {Kachulin}}, \bibinfo {author} {\bibfnamefont {A.}~\bibnamefont {Dyachenko}},
  \ and\ \bibinfo {author} {\bibfnamefont {V.}~\bibnamefont {Zakharov}},\
  }\href {\doibase 10.3390/fluids5020067} {\bibfield  {journal} {\bibinfo
  {journal} {Fluids}\ }\textbf {\bibinfo {volume} {5}},\ \bibinfo {pages} {67}
  (\bibinfo {year} {2020})}\BibitemShut {NoStop}%
\bibitem [{\citenamefont {Costa}\ \emph {et~al.}(2014)\citenamefont {Costa},
  \citenamefont {Osborne}, \citenamefont {Resio}, \citenamefont {Alessio},
  \citenamefont {Chriv\`{\i}}, \citenamefont {Saggese}, \citenamefont
  {Bellomo},\ and\ \citenamefont {Long}}]{Costa:14}%
  \BibitemOpen
  \bibfield  {author} {\bibinfo {author} {\bibfnamefont {A.}~\bibnamefont
  {Costa}}, \bibinfo {author} {\bibfnamefont {A.~R.}\ \bibnamefont {Osborne}},
  \bibinfo {author} {\bibfnamefont {D.~T.}\ \bibnamefont {Resio}}, \bibinfo
  {author} {\bibfnamefont {S.}~\bibnamefont {Alessio}}, \bibinfo {author}
  {\bibfnamefont {E.}~\bibnamefont {Chriv\`{\i}}}, \bibinfo {author}
  {\bibfnamefont {E.}~\bibnamefont {Saggese}}, \bibinfo {author} {\bibfnamefont
  {K.}~\bibnamefont {Bellomo}}, \ and\ \bibinfo {author} {\bibfnamefont
  {C.~E.}\ \bibnamefont {Long}},\ }\href {\doibase
  10.1103/PhysRevLett.113.108501} {\bibfield  {journal} {\bibinfo  {journal}
  {Phys. Rev. Lett.}\ }\textbf {\bibinfo {volume} {113}},\ \bibinfo {pages}
  {108501} (\bibinfo {year} {2014})}\BibitemShut {NoStop}%
\bibitem [{\citenamefont {Perrard}\ \emph {et~al.}(2015)\citenamefont
  {Perrard}, \citenamefont {Deike}, \citenamefont {Duch\^ene},\ and\
  \citenamefont {Pham}}]{Perrard:15}%
  \BibitemOpen
  \bibfield  {author} {\bibinfo {author} {\bibfnamefont {S.}~\bibnamefont
  {Perrard}}, \bibinfo {author} {\bibfnamefont {L.}~\bibnamefont {Deike}},
  \bibinfo {author} {\bibfnamefont {C.}~\bibnamefont {Duch\^ene}}, \ and\
  \bibinfo {author} {\bibfnamefont {C.-T.}\ \bibnamefont {Pham}},\ }\href
  {\doibase 10.1103/PhysRevE.92.011002} {\bibfield  {journal} {\bibinfo
  {journal} {Phys. Rev. E}\ }\textbf {\bibinfo {volume} {92}},\ \bibinfo
  {pages} {011002} (\bibinfo {year} {2015})}\BibitemShut {NoStop}%
\bibitem [{\citenamefont {Redor}\ \emph {et~al.}(2019)\citenamefont {Redor},
  \citenamefont {Barth\'elemy}, \citenamefont {Michallet}, \citenamefont
  {Onorato},\ and\ \citenamefont {Mordant}}]{Redor:19}%
  \BibitemOpen
  \bibfield  {author} {\bibinfo {author} {\bibfnamefont {I.}~\bibnamefont
  {Redor}}, \bibinfo {author} {\bibfnamefont {E.}~\bibnamefont {Barth\'elemy}},
  \bibinfo {author} {\bibfnamefont {H.}~\bibnamefont {Michallet}}, \bibinfo
  {author} {\bibfnamefont {M.}~\bibnamefont {Onorato}}, \ and\ \bibinfo
  {author} {\bibfnamefont {N.}~\bibnamefont {Mordant}},\ }\href {\doibase
  10.1103/PhysRevLett.122.214502} {\bibfield  {journal} {\bibinfo  {journal}
  {Phys. Rev. Lett.}\ }\textbf {\bibinfo {volume} {122}},\ \bibinfo {pages}
  {214502} (\bibinfo {year} {2019})}\BibitemShut {NoStop}%
\bibitem [{\citenamefont {Schwache}\ and\ \citenamefont
  {Mitschke}(1997)}]{Schwache:97}%
  \BibitemOpen
  \bibfield  {author} {\bibinfo {author} {\bibfnamefont {A.}~\bibnamefont
  {Schwache}}\ and\ \bibinfo {author} {\bibfnamefont {F.}~\bibnamefont
  {Mitschke}},\ }\href {\doibase 10.1103/PhysRevE.55.7720} {\bibfield
  {journal} {\bibinfo  {journal} {Phys. Rev. E}\ }\textbf {\bibinfo {volume}
  {55}},\ \bibinfo {pages} {7720} (\bibinfo {year} {1997})}\BibitemShut
  {NoStop}%
\bibitem [{\citenamefont {Marcucci}\ \emph {et~al.}(2019)\citenamefont
  {Marcucci}, \citenamefont {Pierangeli}, \citenamefont {Agranat},
  \citenamefont {Lee}, \citenamefont {DelRe},\ and\ \citenamefont
  {Conti}}]{marcucci_topological_2019}%
  \BibitemOpen
  \bibfield  {author} {\bibinfo {author} {\bibfnamefont {G.}~\bibnamefont
  {Marcucci}}, \bibinfo {author} {\bibfnamefont {D.}~\bibnamefont
  {Pierangeli}}, \bibinfo {author} {\bibfnamefont {A.~J.}\ \bibnamefont
  {Agranat}}, \bibinfo {author} {\bibfnamefont {R.-K.}\ \bibnamefont {Lee}},
  \bibinfo {author} {\bibfnamefont {E.}~\bibnamefont {DelRe}}, \ and\ \bibinfo
  {author} {\bibfnamefont {C.}~\bibnamefont {Conti}},\ }\href@noop {}
  {\bibfield  {journal} {\bibinfo  {journal} {Nature Communications}\ }\textbf
  {\bibinfo {volume} {10}} (\bibinfo {year} {2019})}\BibitemShut {NoStop}%
\bibitem [{\citenamefont {El}\ \emph {et~al.}(2016)\citenamefont {El},
  \citenamefont {Khamis},\ and\ \citenamefont {Tovbis}}]{GEl:16}%
  \BibitemOpen
  \bibfield  {author} {\bibinfo {author} {\bibfnamefont {G.~A.}\ \bibnamefont
  {El}}, \bibinfo {author} {\bibfnamefont {E.~G.}\ \bibnamefont {Khamis}}, \
  and\ \bibinfo {author} {\bibfnamefont {A.}~\bibnamefont {Tovbis}},\
  }\href@noop {} {\bibfield  {journal} {\bibinfo  {journal} {Nonlinearity}\
  }\textbf {\bibinfo {volume} {29}},\ \bibinfo {pages} {2798} (\bibinfo {year}
  {2016})}\BibitemShut {NoStop}%
\bibitem [{\citenamefont {Gelash}\ and\ \citenamefont
  {Agafontsev}(2018)}]{Gelash:18}%
  \BibitemOpen
  \bibfield  {author} {\bibinfo {author} {\bibfnamefont {A.~A.}\ \bibnamefont
  {Gelash}}\ and\ \bibinfo {author} {\bibfnamefont {D.~S.}\ \bibnamefont
  {Agafontsev}},\ }\href {\doibase 10.1103/PhysRevE.98.042210} {\bibfield
  {journal} {\bibinfo  {journal} {Phys. Rev. E}\ }\textbf {\bibinfo {volume}
  {98}},\ \bibinfo {pages} {042210} (\bibinfo {year} {2018})}\BibitemShut
  {NoStop}%
\bibitem [{\citenamefont {Bonnefoy}\ \emph {et~al.}(2020)\citenamefont
  {Bonnefoy}, \citenamefont {Tikan}, \citenamefont {Copie}, \citenamefont
  {Suret}, \citenamefont {Ducrozet}, \citenamefont {Prabhudesai}, \citenamefont
  {Michel}, \citenamefont {Cazaubiel}, \citenamefont {Falcon}, \citenamefont
  {El},\ and\ \citenamefont {Randoux}}]{Bonnefoy:20}%
  \BibitemOpen
  \bibfield  {author} {\bibinfo {author} {\bibfnamefont {F.}~\bibnamefont
  {Bonnefoy}}, \bibinfo {author} {\bibfnamefont {A.}~\bibnamefont {Tikan}},
  \bibinfo {author} {\bibfnamefont {F.}~\bibnamefont {Copie}}, \bibinfo
  {author} {\bibfnamefont {P.}~\bibnamefont {Suret}}, \bibinfo {author}
  {\bibfnamefont {G.}~\bibnamefont {Ducrozet}}, \bibinfo {author}
  {\bibfnamefont {G.}~\bibnamefont {Prabhudesai}}, \bibinfo {author}
  {\bibfnamefont {G.}~\bibnamefont {Michel}}, \bibinfo {author} {\bibfnamefont
  {A.}~\bibnamefont {Cazaubiel}}, \bibinfo {author} {\bibfnamefont
  {E.}~\bibnamefont {Falcon}}, \bibinfo {author} {\bibfnamefont
  {G.}~\bibnamefont {El}}, \ and\ \bibinfo {author} {\bibfnamefont
  {S.}~\bibnamefont {Randoux}},\ }\href {\doibase
  10.1103/PhysRevFluids.5.034802} {\bibfield  {journal} {\bibinfo  {journal}
  {Phys. Rev. Fluids}\ }\textbf {\bibinfo {volume} {5}},\ \bibinfo {pages}
  {034802} (\bibinfo {year} {2020})}\BibitemShut {NoStop}%
\bibitem [{\citenamefont {Osborne}(2010)}]{osborne2010nonlinear}%
  \BibitemOpen
  \bibfield  {author} {\bibinfo {author} {\bibfnamefont {A.}~\bibnamefont
  {Osborne}},\ }\href@noop {} {\emph {\bibinfo {title} {{Nonlinear ocean
  waves}}}}\ (\bibinfo  {publisher} {Academic Press},\ \bibinfo {year}
  {2010})\BibitemShut {NoStop}%
\bibitem [{\citenamefont {Gelash}\ \emph {et~al.}(2019)\citenamefont {Gelash},
  \citenamefont {Agafontsev}, \citenamefont {Zakharov}, \citenamefont {El},
  \citenamefont {Randoux},\ and\ \citenamefont {Suret}}]{Gelash:19}%
  \BibitemOpen
  \bibfield  {author} {\bibinfo {author} {\bibfnamefont {A.}~\bibnamefont
  {Gelash}}, \bibinfo {author} {\bibfnamefont {D.}~\bibnamefont {Agafontsev}},
  \bibinfo {author} {\bibfnamefont {V.}~\bibnamefont {Zakharov}}, \bibinfo
  {author} {\bibfnamefont {G.}~\bibnamefont {El}}, \bibinfo {author}
  {\bibfnamefont {S.}~\bibnamefont {Randoux}}, \ and\ \bibinfo {author}
  {\bibfnamefont {P.}~\bibnamefont {Suret}},\ }\href {\doibase
  10.1103/PhysRevLett.123.234102} {\bibfield  {journal} {\bibinfo  {journal}
  {Phys. Rev. Lett.}\ }\textbf {\bibinfo {volume} {123}},\ \bibinfo {pages}
    {234102} (\bibinfo {year} {2019})}\BibitemShut {NoStop}%

\bibitem{note_suppl} see Supplemental Material, which 
  includes ref. \cite{Zakharov:72,Randoux:16a,Goullet:11,Dommermuth_1987,West_1987,Ducrozet_2012,CODE,Ducrozet_2012,Bonnefoy_2010}, for numerical simulations of the water wave
  experiments together with a description of the numerical methods
  used for nonlinear spectral analysis and synthesis of the wavefields.
  
\bibitem [{\citenamefont {Chekhovskoy}\ \emph {et~al.}(2019)\citenamefont
  {Chekhovskoy}, \citenamefont {Shtyrina}, \citenamefont {Fedoruk},
  \citenamefont {Medvedev},\ and\ \citenamefont {Turitsyn}}]{Chekhovskoy:19}%
  \BibitemOpen
  \bibfield  {author} {\bibinfo {author} {\bibfnamefont {I.~S.}\ \bibnamefont
  {Chekhovskoy}}, \bibinfo {author} {\bibfnamefont {O.~V.}\ \bibnamefont
  {Shtyrina}}, \bibinfo {author} {\bibfnamefont {M.~P.}\ \bibnamefont
  {Fedoruk}}, \bibinfo {author} {\bibfnamefont {S.~B.}\ \bibnamefont
  {Medvedev}}, \ and\ \bibinfo {author} {\bibfnamefont {S.~K.}\ \bibnamefont
  {Turitsyn}},\ }\href {\doibase 10.1103/PhysRevLett.122.153901} {\bibfield
  {journal} {\bibinfo  {journal} {Phys. Rev. Lett.}\ }\textbf {\bibinfo
  {volume} {122}},\ \bibinfo {pages} {153901} (\bibinfo {year}
  {2019})}\BibitemShut {NoStop}%
\bibitem [{\citenamefont {Randoux}\ \emph {et~al.}(2018)\citenamefont
  {Randoux}, \citenamefont {Suret}, \citenamefont {Chabchoub}, \citenamefont
  {Kibler},\ and\ \citenamefont {El}}]{Randoux:18}%
  \BibitemOpen
  \bibfield  {author} {\bibinfo {author} {\bibfnamefont {S.}~\bibnamefont
  {Randoux}}, \bibinfo {author} {\bibfnamefont {P.}~\bibnamefont {Suret}},
  \bibinfo {author} {\bibfnamefont {A.}~\bibnamefont {Chabchoub}}, \bibinfo
  {author} {\bibfnamefont {B.}~\bibnamefont {Kibler}}, \ and\ \bibinfo {author}
  {\bibfnamefont {G.}~\bibnamefont {El}},\ }\href {\doibase
  10.1103/PhysRevE.98.022219} {\bibfield  {journal} {\bibinfo  {journal} {Phys.
  Rev. E}\ }\textbf {\bibinfo {volume} {98}},\ \bibinfo {pages} {022219}
  (\bibinfo {year} {2018})}\BibitemShut {NoStop}%
\bibitem [{\citenamefont {Zakharov}\ and\ \citenamefont
  {Shabat}(1972)}]{Zakharov:72}%
  \BibitemOpen
  \bibfield  {author} {\bibinfo {author} {\bibfnamefont {V.~E.}\ \bibnamefont
  {Zakharov}}\ and\ \bibinfo {author} {\bibfnamefont {A.~B.}\ \bibnamefont
  {Shabat}},\ }\href@noop {} {\bibfield  {journal} {\bibinfo  {journal} {Sov.
  Phys.--JETP}\ }\textbf {\bibinfo {volume} {34}},\ \bibinfo {pages} {62}
  (\bibinfo {year} {1972})}\BibitemShut {NoStop}%
\bibitem [{\citenamefont {Randoux}\ \emph {et~al.}(2016)\citenamefont
  {Randoux}, \citenamefont {Suret},\ and\ \citenamefont {El}}]{Randoux:16a}%
  \BibitemOpen
  \bibfield  {author} {\bibinfo {author} {\bibfnamefont {S.}~\bibnamefont
  {Randoux}}, \bibinfo {author} {\bibfnamefont {P.}~\bibnamefont {Suret}}, \
  and\ \bibinfo {author} {\bibfnamefont {G.}~\bibnamefont {El}},\ }\href@noop
  {} {\bibfield  {journal} {\bibinfo  {journal} {Scientific reports}\ }\textbf
  {\bibinfo {volume} {6}},\ \bibinfo {pages} {29238} (\bibinfo {year}
  {2016})}\BibitemShut {NoStop}%
\bibitem [{\citenamefont {Goullet}\ and\ \citenamefont
  {Choi}(2011)}]{Goullet:11}%
  \BibitemOpen
  \bibfield  {author} {\bibinfo {author} {\bibfnamefont {A.}~\bibnamefont
  {Goullet}}\ and\ \bibinfo {author} {\bibfnamefont {W.}~\bibnamefont {Choi}},\
  }\href {\doibase 10.1063/1.3533961} {\bibfield  {journal} {\bibinfo
  {journal} {Physics of Fluids}\ }\textbf {\bibinfo {volume} {23}},\ \bibinfo
  {pages} {016601} (\bibinfo {year} {2011})}\BibitemShut {NoStop}%
\bibitem [{\citenamefont {Dommermuth}\ and\ \citenamefont
  {Yue}(1987)}]{Dommermuth_1987}%
  \BibitemOpen
  \bibfield  {author} {\bibinfo {author} {\bibfnamefont {D.~G.}\ \bibnamefont
  {Dommermuth}}\ and\ \bibinfo {author} {\bibfnamefont {D.~K.}\ \bibnamefont
  {Yue}},\ }\href@noop {} {\bibfield  {journal} {\bibinfo  {journal} {J. Fluid
  Mech.}\ }\textbf {\bibinfo {volume} {184}},\ \bibinfo {pages} {267} (\bibinfo
  {year} {1987})}\BibitemShut {NoStop}%
\bibitem [{\citenamefont {West}\ \emph {et~al.}(1987)\citenamefont {West},
  \citenamefont {Brueckner}, \citenamefont {Janda}, \citenamefont {Milder},\
  and\ \citenamefont {Milton}}]{West_1987}%
  \BibitemOpen
  \bibfield  {author} {\bibinfo {author} {\bibfnamefont {B.~J.}\ \bibnamefont
  {West}}, \bibinfo {author} {\bibfnamefont {K.~A.}\ \bibnamefont {Brueckner}},
  \bibinfo {author} {\bibfnamefont {R.~S.}\ \bibnamefont {Janda}}, \bibinfo
  {author} {\bibfnamefont {D.~M.}\ \bibnamefont {Milder}}, \ and\ \bibinfo
  {author} {\bibfnamefont {R.~L.}\ \bibnamefont {Milton}},\ }\href@noop {}
  {\bibfield  {journal} {\bibinfo  {journal} {J. Geophys. Res.}\ }\textbf
  {\bibinfo {volume} {92}},\ \bibinfo {pages} {11803} (\bibinfo {year}
  {1987})}\BibitemShut {NoStop}%
\bibitem [{\citenamefont {Ducrozet}\ \emph {et~al.}(2012)\citenamefont
  {Ducrozet}, \citenamefont {Bonnefoy}, \citenamefont {Touzé},\ and\
  \citenamefont {Ferrant}}]{Ducrozet_2012}%
  \BibitemOpen
  \bibfield  {author} {\bibinfo {author} {\bibfnamefont {G.}~\bibnamefont
  {Ducrozet}}, \bibinfo {author} {\bibfnamefont {F.}~\bibnamefont {Bonnefoy}},
  \bibinfo {author} {\bibfnamefont {D.~L.}\ \bibnamefont {Touzé}}, \ and\
  \bibinfo {author} {\bibfnamefont {P.}~\bibnamefont {Ferrant}},\ }\href@noop
  {} {\bibfield  {journal} {\bibinfo  {journal} {Eur. J. Mech. B. Fluids}\
  }\textbf {\bibinfo {volume} {34}},\ \bibinfo {pages} {19} (\bibinfo {year}
  {2012})}\BibitemShut {NoStop}%
\bibitem [{COD()}]{CODE}%
  \BibitemOpen
  \href@noop {} {}\bibinfo {note} {Ecole Centrale Nantes, LHEEA, Open-source
  release of HOS-NWT, https://github.com/LHEEA/HOS-NWT}\BibitemShut {NoStop}%
\bibitem [{\citenamefont {Bonnefoy}\ \emph {et~al.}(2010)\citenamefont
  {Bonnefoy}, \citenamefont {Ducrozet}, \citenamefont {Touzé},\ and\
  \citenamefont {Ferrant}}]{Bonnefoy_2010}%
  \BibitemOpen
  \bibfield  {author} {\bibinfo {author} {\bibfnamefont {F.}~\bibnamefont
  {Bonnefoy}}, \bibinfo {author} {\bibfnamefont {G.}~\bibnamefont {Ducrozet}},
  \bibinfo {author} {\bibfnamefont {D.~L.}\ \bibnamefont {Touzé}}, \ and\
  \bibinfo {author} {\bibfnamefont {P.}~\bibnamefont {Ferrant}},\ }\enquote
  {\bibinfo {title} {Time domain simulation of nonlinear water waves using
  spectral methods},}\ in\ \href@noop {} {\emph {\bibinfo {booktitle} {Advances
  in Numerical Simulation of Nonlinear Water Waves}}}\ (\bibinfo {year}
  {2010})\ pp.\ \bibinfo {pages} {129--164}\BibitemShut {NoStop}%
\end{thebibliography}
%

\pagebreak
\renewcommand{\theequation}{S\arabic{equation}}
\setcounter{equation}{0}
\onecolumngrid

\maketitle

\renewcommand{\theequation}{S\arabic{equation}}
\renewcommand{\thefigure}{S\arabic{figure}}

\begin{center}
{\bf Supplemental Material for ``Nonlinear spectral synthesis of soliton gases in deep-water surface gravity waves''}\\
\end{center}

\begin{center}
  Pierre Suret,$^1$ Alexey Tikan,$^1$ F\'elicien Bonnefoy,$^2$ Fran\c{c}ois Copie,$^1$ Guillaume Ducrozet,$^2$ Andrey Gelash,$^{3,4}$ Gaurav Prabhudesai,$^5$ Guillaume Michel,$^6$ Annette Cazaubiel,$^7$ Eric Falcon,$^7$ Gennady El,$^8$ St\'ephane Randoux$^1$
\end{center}

\begin{center}
  {\it $^1$ Univ. Lille, CNRS, UMR 8523 - PhLAM -Physique des Lasers Atomes et Mol\'ecules, F-59 000 Lille, France
  
  $^2$ \'Ecole Centrale de Nantes, LHEEA, UMR 6598 CNRS, F-44 321 Nantes, France
  
  $^3$ Institute of Automation and Electrometry SB RAS, Novosibirsk 630090, Russia
  
  $^4$ Skolkovo Institute of Science and Technology, Moscow 121205, Russia
  
  $^5$ Laboratoire de Physique de l'Ecole normale sup\'erieure, ENS,Universit\'e PSL, CNRS, Sorbonne Universit\'e, Universit\'e Paris-Diderot, Paris, France
  
  $^6$ Sorbonne Universit\'e, CNRS, UMR 7190, Institut Jean Le Rond d’Alembert, F-75 005 Paris, France
  
  $^7$ Universit\'e de Paris, Universit\'e Paris Diderot, MSC, UMR 7057 CNRS, F-75 013 Paris, France
  
  $8$ Department of Mathematics, Physics and Electrical Engineering, Northumbria University, Newcastle upon Tyne, NE1 8ST, United Kingdom
  }
\end{center}

%\date{\today}

The purpose of this Supplemental Material is to provide some
mathematical, numerical and experimental details that are utilized in the Letter.  
All equation, figure, reference numbers
within this document are prepended with ``S'' to distinguish them from
corresponding numbers in the Letter.\\

\section{Nonlinear spectral synthesis of N-soliton solutions of the focusing 1D-NLSE}\label{nss}

In this section, we briefly describe the methodology used for the nonlinear synthesis
of the large soliton ensembles propagating in the one-dimensional water tank. 
More theoretical details can be found in ref. \cite{Gelash:18}. 

We consider the focusing 1D-NLSE in the form 
\begin{equation}\label{NLSE}
 i \psi_t + \frac{1}{2} \psi_{xx}+  |\psi|^2 \psi=0 \, ,
\end{equation}
where $\psi(x,t)$ is a complex wave envelope varying in space $x$ and
time $t$. In the IST method, the NLSE is represented as the
compatibility condition of two linear equations \cite{Zakharov:72},
\begin{equation}\label{LP1}
\Phi_x=
 \begin{pmatrix}
-i \lambda & \psi \\ -\psi^* & i \lambda \\
\end{pmatrix} 
\Phi,
\end{equation}
\begin{equation}\label{LP2}
\Phi_t=
 \begin{pmatrix}
-i \lambda^2+\frac{i}{2}|\psi|^2 & \frac{i}{2} \psi_x +2\lambda \psi \\ \frac{i}{2} \psi_x^* - \lambda \psi^*
&  i \lambda^2 - \frac{i}{2} |\psi|^2 \\
\end{pmatrix} 
\Phi,
\end{equation}
where $\lambda$ is a complex spectral parameter and $\Phi(t,x,\lambda)$ is a
$2 \times 2$ matrix wave function. 

For spatially localized potentials $\psi$ such that $\psi(x,t) \rightarrow 0$
as $|x| \rightarrow \infty$, the eigenvalues $\lambda$ are
presented by a finite number of discrete points with $\Im(\lambda) \ne 0$ 
(discrete spectrum) and the real line $\lambda \in \mathbb{R}$ (continuous spectrum).
The scattering data consists of discrete eigenvalues $\lambda_n$,
$n = 1, ..., N$ , norming constantss $C_n$ for each $\lambda_n$ and the
so-called reflection coefficient $r (\xi)$,
\begin{equation}
\{ r (\xi) ; \, \lambda_n, \, C_n \}
\end{equation}  
where $\xi$ means $\lambda$ on the real axis. 

The simplest reflectionless ($r (\xi)=0$) solution of Eq. (\ref {NLSE}) is
the fundamental soliton which is parametrized by one discrete complex eigenvalue
$\lambda_1$ and one associated complex norming constant $C_1$ that read
\begin{equation}
\lambda_1=-v_1/2+i \, a_1/2,  \quad  C_1 = \exp i(\theta_1 + 2 \lambda_1 x_1)
\end{equation}  
With this setting the one-soliton solution of Eq. (\ref {NLSE}) reads
\begin{equation}
  \psi_{(1)}(x,t)=\frac{a_1 \exp{ \left[ i v_1(x - x_1) + \frac{i}{2} (a_1^2 - v_1^2) t + i \theta_1 \right]}  }{\cosh(a_1(x-x_1)-a_1 v_1 t)} 
\end{equation}  
where $a_1>0$ represents the maximum amplitude of the soliton which moves with the group
velocity $v_1$ in the $(x,t)$ plane. $x_1$ and $\theta_1$ represent the position
and the phase of the soliton at $t=0$, respectively. 

A special class of
solutions of Eq. (\ref {NLSE}), the N-soliton solution (N-SS),
exhibits only a discrete spectrum ($r (\xi)=0$) consisting of N complex-valued
eigenvalues $\lambda_n$, $n=1,..., N$ and their associated norming constants 
$C_n=|C_n|e^{i \phi_n}$. To construct a N-SS at the initial time $t=0$, we
first generate an ensemble of $N$ discrete eigenvalues $\lambda_n=-v_n/2+i \, a_n/2$
and of their associated norming constants $ C_n = \exp i(\theta_n + 2 \lambda_n x_n)$. 
As discussed in details in ref. \cite{Gelash:18}, the generation of the
N-SS is achieved via a recurrent dressing procedure where discrete eigenvalues are
iteratively added starting from the trivial solution $\psi_{(0)}=0$ of Eq. (\ref{NLSE}). 

The recurrence formula used to compute the N-SS is \cite{Gelash:18}
\begin{equation}
  \psi_{(n)}(x,0)=\psi_{(n-1)}(x,0)+2i(\lambda_n - \lambda_n^* ) \frac{q_{n1}^* q_{n2}}{|q_n|^2},
\end{equation}  
where the vector ${\bf q_n}=(q_{n1},q_{n2})^T$ is determined from ${\bf \Phi}_{n-1}$ and the
scattering data of the ${\it n}$th soliton $\{\lambda_n, \, C_n \}$, 
${\bf q_n}(x)={\bf \Phi}_{n-1}^* (x,\lambda_n^*)  \left[\begin{array}{@{}c@{}} 1 \\    C_n \\  \end{array} \right]$. 
The corresponding solution ${\bf \Phi}_{n}(x,\lambda)$ of the Zakharov-Shabat system (\ref{LP1})
is calculated using ${\bf \Phi}_{n-1}$ and the so-called dressing matrix $\bm{\chi}$, 
\begin{equation}
  \bm{\Phi}_{(n)}(x,\lambda)=\bm{\chi}(x,\lambda) . \bm{\Phi}_{(n-1)}(x,\lambda)
\end{equation}  
\begin{equation}
  \bm{\chi}_{ml}(x,\lambda)= \delta_{ml} + \frac{\lambda_n - \lambda_n^*}{\lambda - \lambda_n^*}
  \frac{q_{nm}^* q_{nl}}{|\bm{q_n}|^2}
\end{equation}  
where $m,l=1,2$ and $\delta_{ml}$ is the Kronecker $\delta$ symbol \cite{Gelash:18}. 

\section{Inverse scattering transform analysis of the experimental data}\label{ist}

In this Section, we describe briefly the method used to compute the discrete
IST spectrum from the signals recorded in the water wave experiment. 

The first step for performing the nonlinear analysis of the signals
(water elevation given by $\eta(Z,T)=Re \left[ A(Z,T)e^{i(k_0 Z -\omega_0 T)} \right]$)
recorded in the experiments consists in determining the complex envelope $A(Z,T)$ of the wavefield. 
This is achieved by using standard techniques based on the Hilbert
transform, as discussed e. g. in ref. \cite{osborne2010nonlinear}.
Then, physical quantities are put to dimensionless form using the
connection between physical and dimensionless variables that are
provided in the Letter and recalled here for the sake of simplicity: $\psi=A/\sqrt{<|A_0(T)|^2>}$, $T=Z/L_{NL}$, $x=(T-Z/C_g) \sqrt{g/(2 L_{NL})}$
with the nonlinear length being defined as $L_{NL}=1/(\alpha k_0^3 <|A_0(T)|^2>)$. 
The brackets denote average over time. 
Finally the IST discrete spectrum is determined by solving the Zakharov-Shabat
system (\ref{LP1}) using the Fourier collocation method and following a procedure
used and described in ref. \cite{yang2010nonlinear, Randoux:16a, Randoux:18,Bonnefoy:20}

\section{Integrable versus non-integrable dynamics in the ensemble of 16 solitons}\label{dysthe_16}

In this Section, we use numerical simulations of the focusing 1D-NLSE 
and of a modified (non-integrable) 1D-NLSE to show the role 
of higher order effects on the observed space-time dynamics and on the spectral (IST)
features that characterize the evolution of the ensemble of $16$ solitons
considered in Fig. 1 of the Letter. 

Following the work reported in ref. \cite{Goullet:11}, higher-order effects in 1D
water wave experiments can be described by a modified NLSE written under the form
of a spatial evolution equation
\begin{equation}\label{dysthe_eq}
%  \begin{split}
  \frac{\partial A}{\partial Z} = i \frac{k_0}{\omega_0^2}
  \frac{\partial^2 A}{\partial T^2} + i \alpha k_0^3 |A|^2 A
  - \frac{k_0^3}{\omega_0} \left( 6|A|^2 \frac{\partial A}{\partial T}
  +2A\frac{\partial |A|^2}{\partial T}
  -2 i A \mathcal{H}\left[\frac{\partial |A|^2}{\partial T } \right] \right)
  ,
%  \end{split}
\end{equation}
where $A(Z,T)$ represents the complex envelope of the wave field
and $\mathcal{H}$ is the Hilbert transform defined by
$\mathcal{H}[f]=(1/\pi) \int_{-\infty}^{+\infty} f(\xi)/(\xi-T)d\xi$.
When the last three terms are neglected in Eq. (\ref{dysthe_eq}),
the integrable 1D-NLSE is recovered.

Neglecting the last three terms in Eq. (\ref{dysthe_eq}),
Fig. S1 shows results obtained from the numerical simulation of Eq. (\ref{dysthe_eq})
(integrable focusing 1D-NLSE) for the ensemble of $16$ solitons considered in the experiments reported
in Fig. 1 of the Letter. Fig. S1(a)  (resp. Fig. S1(b)) shows the modulus $|A(Z,T)|$ of the wavefield
that is computed at $Z_1=6$ m (resp. at $Z_{20}=120$ m). Despite the significantly
complicated space time evolution shown in Fig. S1(e) over the $120$ m-long propagation
distance, the dynamics is integrable which implies that the discrete IST spectrum
remains perfectly unchanged between $Z_1$ and $Z_{20}$, compare Fig. S1(c)
and Fig.S1(d). \\

\begin{figure}[H]
  \centering  
  \includegraphics[width=1\textwidth]{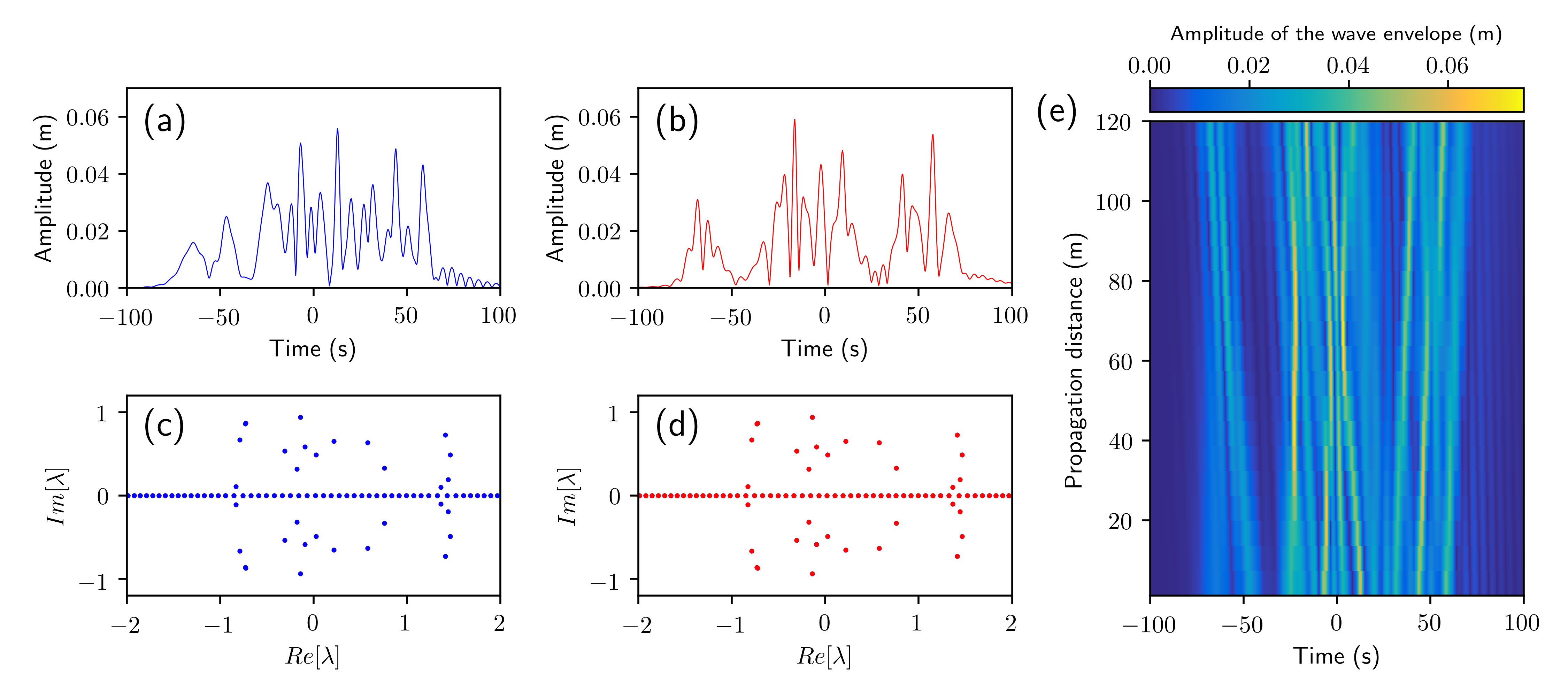}
  \caption{Integrable dynamics. Numerical simulations of the integrable focusing 1D-NLSE
    (Eq. (\ref{dysthe_eq}) where the last three terms are neglected)
    for the ensemble of $16$ solitons considered in Fig. 1 of the Letter. 
    (a) Modulus $|A(Z_1,T)|$ of the wave envelope at $Z_1=6$ m and (c)
    corresponding discrete IST spectrum.
     (b) Modulus $|A(Z_{20},t)|$ of the wave envelope at $Z_{20}=120$ m and (d)
    corresponding discrete IST spectrum.
    (e) Space-time plot showing the nonlinear evolution of the modulus $|A(Z,T)|$
  of the wave envelope.}
\end{figure}

\begin{figure}[H]
  \centering
  \includegraphics[width=1\textwidth]{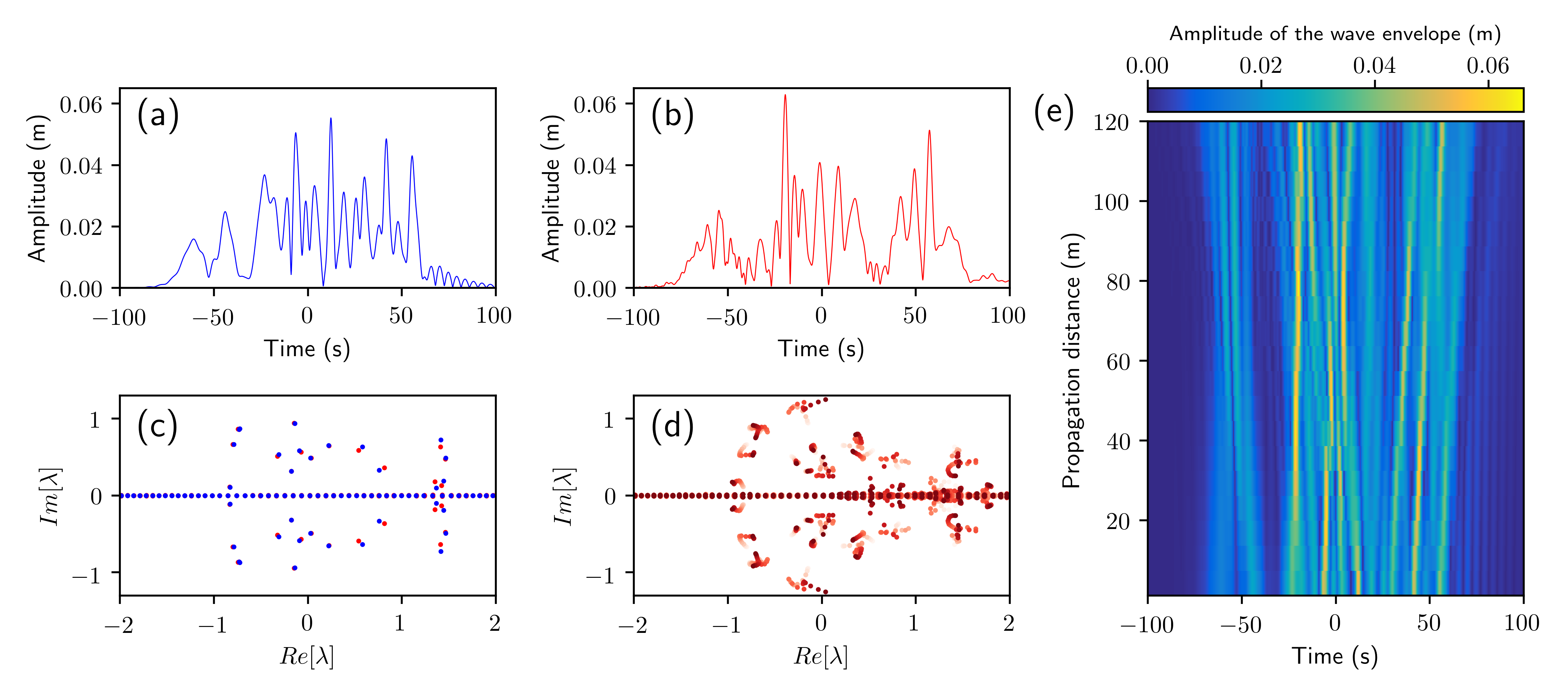}
  \caption{Non-integrable dynamics. Numerical simulations of 
    Eq. (\ref{dysthe_eq}) 
    for the ensemble of $16$ solitons considered in Fig. 1 of the Letter. 
    (a) Modulus $|A(Z_1,t)|$ of the wave envelope at $Z_1=6$ m and (c)
    corresponding discrete IST spectrum.
     (b) Modulus $|A(Z_{20},t)|$ of the wave envelope at $Z_{20}=120$ m . (d)
    Space evolution of the discrete IST spectra along the tank from $Z_1 = 6$ m (light red) to $Z_{20} = 120$ m (dark red).
    (e) Space-time plot showing the nonlinear evolution of the modulus $|A(Z,T)|$
  of the wave envelope.
}
\end{figure}

If higher order effects described by the three last terms in Eq. (\ref{dysthe_eq})
are taken into account, the space-time evolution is slightly perturbed compared
to the integrable case, compare Fig. S2(b) with Fig. S1(b) and Fig. S2(e) with Fig. S1(e).
Contrary to results reported in Fig. S1, the isospectrality condition is now
not verified because of the higher-order effects that break the integrability of
the wave dynamics. Fig. S2(d) shows clearly that
each of the $16$ discrete eigenvalues composing the random wavefield
does not remain invariant over propagation distance but follows an individual
trajectory in the complex plane, as already e.g. evidenced in ref. \cite{Chekhovskoy:19}
in numerical simulations of a laser system. Similar spectral (IST) results 
have been presented in experimental results reported in Fig. 1 of the Letter.
However  clean trajectories in the complex IST plane
cannot be clearly identified in experiments because of small calibration
errors in the measurement of the wave elevation. 

\section{Influence of higher-order efects on the gas of $128$ solitons}\label{dysthe_128}

In this Section, we show that Eq. (\ref{dysthe_eq}) describes  well
dynamical and statistical features reported in Fig. 2 and in Fig. 3 of the Letter
for the gas of $128$ solitons. \\

Fig. S3 shows numerical simulations of Eq. (\ref{dysthe_eq}) that are made
with physical values characterizing the experiments presented in the Letter for the gas
of $128$ solitons. Dynamical and spectral features very similar to those reported in
Fig. 2 of the Letter are found in numerical simulations reported in Fig. S3. 
In particular the isospectrally condition is not verified because of the
perturbative higher order effects described by the last three terms in Eq. (\ref{dysthe_eq}).
This results in discrete IST spectra that significantly change with propagation
distance (compare Fig. S3(c) and Fig. S3(d)) even though they remain confined
in a well defined region in the complex plane. 

Fig. S4 shows the normalized density of states, i.e.
the probability density function
of the complex-valued discrete eigenvalues characterizing
the SG over the time interval $\Delta T = 1200$ s
It is determined at different propagation distances from numerical
simulations of Eq. (\ref{dysthe_eq}). 

\begin{figure}[H]
  \centering    
  \includegraphics[width=1\textwidth]{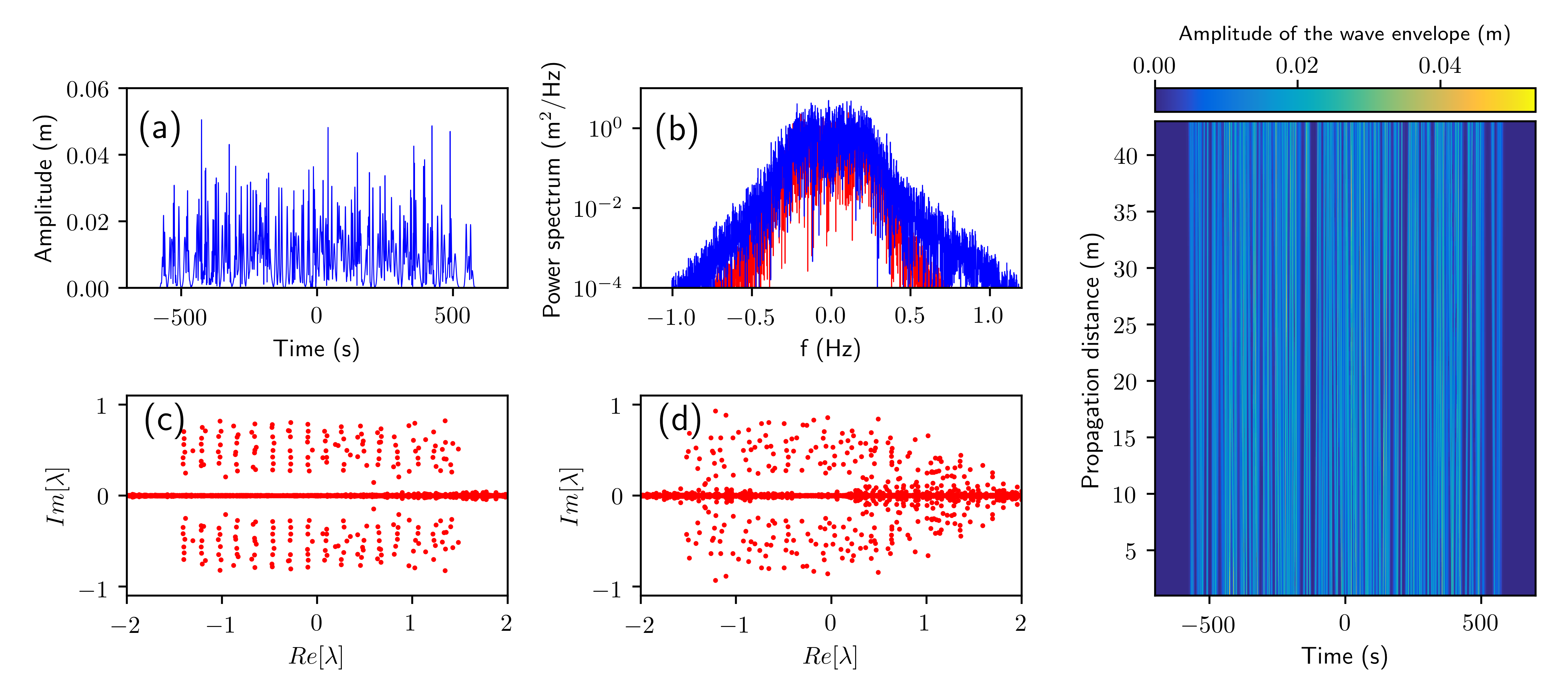}
  \caption{Numerical simlations of Eq. (\ref{dysthe_eq}).
    Gas of $N = 128$ solitons propagating in the 1D water tank. (a) 
Modulus of the wave envelope at $Z_1 = 6$ m. (b) Fourier power spectra of
wave elevation at $Z_1 = 6$ m (red line) and at $Z_{20} = 120$ m (blue line).
(c) Discrete IST spectrum measured at $Z_1 = 6$ m. (d)
Discrete IST spectrum measured at $Z_{20} = 120$ m. (e)
Space-time evolution of modulus of the wave envelope. 
Parameters of the simulation are $f_0 = 1.15$ Hz,
$k_0=5.32$ m$^{-1}$, $L_{NL}=45$ m ($<|A_0(T)|^2>=1.58 \,\times 10^{-4}$ m$^{2}$.}
\end{figure}

\begin{figure}[H]
    \centering    
  \includegraphics[width=1\textwidth]{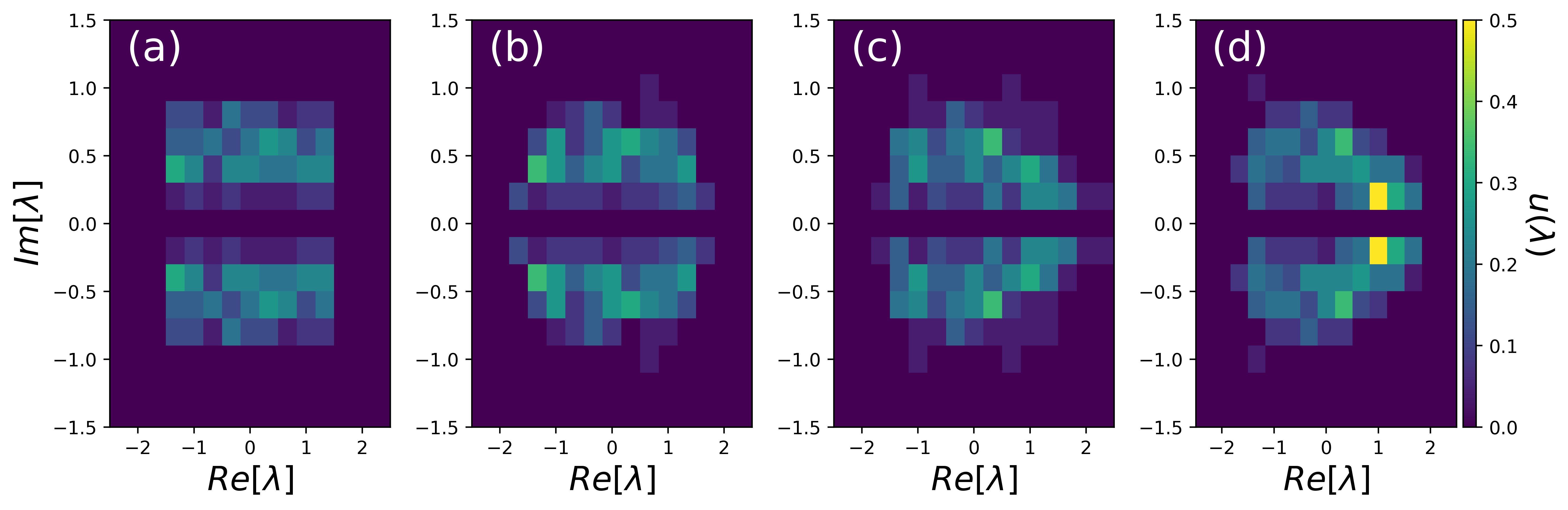}
  \caption{Numerical simlations of Eq. (\ref{dysthe_eq}).
    Statistical analysis of discrete IST spectra of Fig. S3
     showing the slow evolution of the
    DOS $u(\lambda)$ (the probability density function of the discrete IST eigenvalues in the
    complex plane) as a function of propagation distance in the water tank: 
    (a) $Z_1=6$ m, (b) $Z_3=18$ m, (c) $Z_{10}=60$ m, (d) $Z_{20}=120$ m.
Parameters of the simulation are $f_0 = 1.15$ Hz,
$k_0=5.32$ m$^{-1}$, $L_{NL}=45$ m ($<|A_0(T)|^2>=1.58 \,\times 10^{-4}$ m$^{2}$.
}
\end{figure}

\section{Direct numerical simulations of Euler's equations for the gas of $128$ solitons}\label{hos_128}

Direct numerical simulations of the Euler's equations have been
performed with the efficient and accurate High-Order Spectral (HOS)
method \cite{Dommermuth_1987,West_1987}.
The numerical model used in our numerical simulations  reproduce the main
features of the water tank, namely: i) the generation of waves through
a wave maker and ii) the absorption of reflected waves with an
absorbing beach. To this end a Numerical Wave Tank, entitled HOS-NWT,
has been developed \cite{Ducrozet_2012} (the code being available
open-source \cite{CODE}). It uses the exact same wave maker’s motions
than in the experiments for a simplified comparison procedure. This
specific model has been widely validated in different configurations
and more details can be found in \cite{Ducrozet_2012,Bonnefoy_2010}.

\begin{figure}[H]
    \centering    
  \includegraphics[width=1\textwidth]{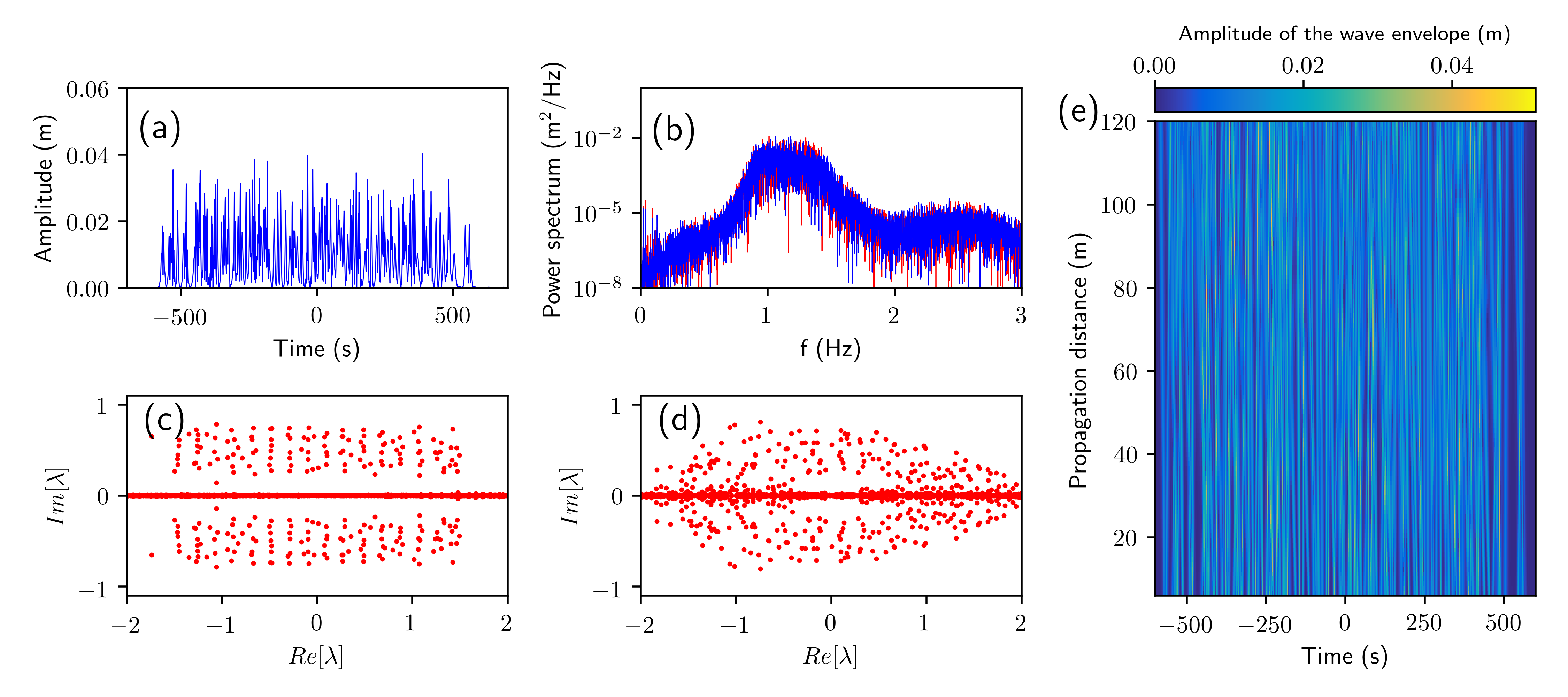}
  \caption{Numerical
    simulations of the Euler's equations corresponding
  to the experimental results shown in Fig. 2 of the manuscript ($N=128$).
    (a) Modulus of the wave envelope 
    at $Z_1=6$ m, close to the wavemaker. (b) Fourier power spectra
    of wave elevation at $Z_1=6$ m (blue line) and at $Z_{20}=120$ m (red line).
    (c) Discrete IST spectrum measured at $Z_1=6$ m.
    (d) Discrete IST spectrum measured at $Z_{20}=120$ m.      
    (e) Space-time evolution of modulus of the
    wave envelope.
    Parameters of the simulation are $f_0 = 1.15$ Hz,
$k_0=5.32$ m$^{-1}$, $L_{NL}=45$ m ($<|A_0(T)|^2>=1.58 \,\times 10^{-4}$ m$^{2}$.
    }
\end{figure}

Fig. S5 shows numerical simulations of Euler's equations that are made
with physical values characterizing the experiments presented in the Letter for the gas
of $128$ solitons. Dynamical and spectral features very similar to those reported in
Fig. 2 of the Letter are found in numerical simulations reported in Fig. S4. 

Fig. S6 shows the normalized density of states, i.e.
the probability density function of the complex-valued discrete
eigenvalues characterizing the SG.
It is determined at different propagation distances from numerical
simulations of Euler's equations \cite{Dommermuth_1987,West_1987}.

\begin{figure}[H]
    \centering    
  \includegraphics[width=1\textwidth]{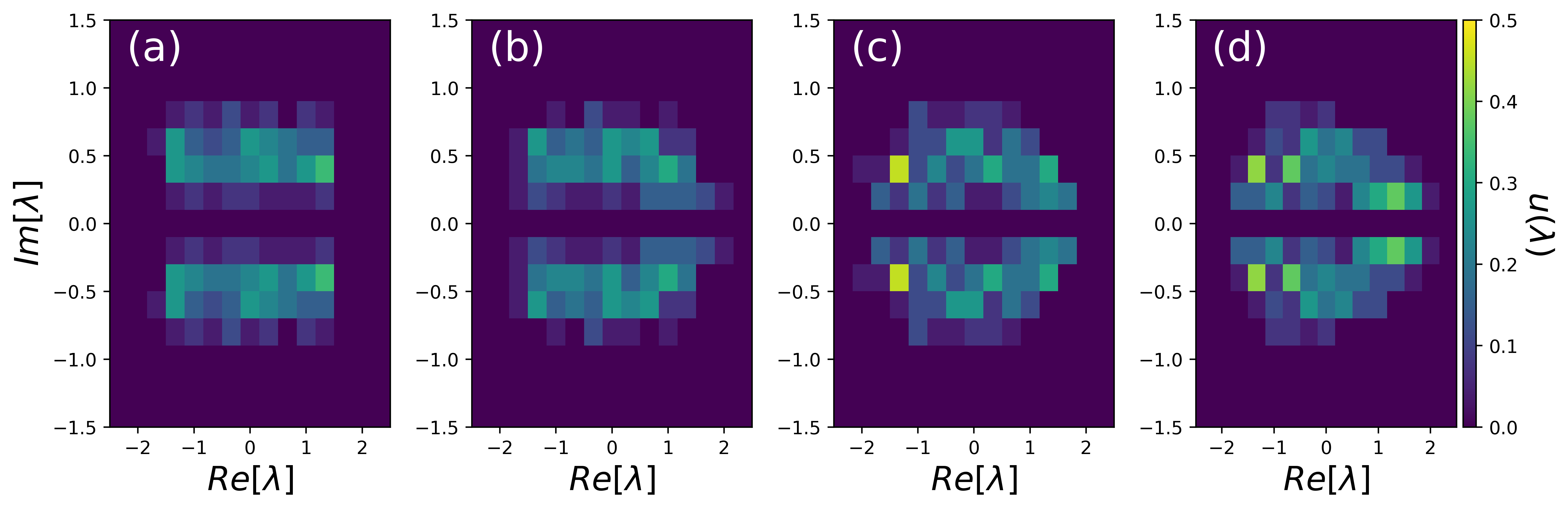}
  \caption{Numerical simlations of Eq. (\ref{dysthe_eq}).
    Statistical analysis of discrete IST spectra of Fig. S5
     showing the slow evolution of the
    DOS $u(\lambda)$ (the probability density function of the discrete IST eigenvalues in the
    complex plane) as a function of propagation distance in the water tank: 
    (a) $Z_1=6$ m, (b) $Z_3=18$ m, (c) $Z_{10}=60$ m, (d) $Z_{20}=120$ m.
Parameters of the simulation are $f_0 = 1.15$ Hz,
$k_0=5.32$ m$^{-1}$, $L_{NL}=45$ m ($<|A_0(T)|^2>=1.58 \,\times 10^{-4}$ m$^{2}$.
}
\end{figure}

\end{document}